\documentclass[preprint,12pt]{elsarticle}
\usepackage[latin9]{inputenc}
\PassOptionsToPackage{linesnumbered, ruled, vlined}{algorithm2e}
\usepackage{array}
\usepackage{mathtools}
\usepackage{multirow}
\usepackage{algorithm2e}
\usepackage{amssymb}
\usepackage{graphicx}
\usepackage{booktabs}
\usepackage{amsthm}
\usepackage{enumitem}
\usepackage{xcolor}
\usepackage{float}
\usepackage{soul}
\theoremstyle{definition}
\newtheorem{definition}{Definition}[section]
\usepackage[a4paper, total={6in, 10in}]{geometry}
\makeatletter
\providecommand{\tabularnewline}{\\}
\usepackage{lineno}
\journal{IJSEKE (Accepted)}
\makeatother
\begin{document}
\begin{frontmatter}

\title{\textbf{Generation and Application of Constrained Interaction Test Suites Using Base Forbidden Tuples With Mixed Neighborhood Tabu Search}}

\author{\textbf{Imad H. Hasan}}

\address{Software and Informatics Engineering, Salahaddin University-Erbil (SUE)\\ 44002, Erbil, Iraq
\\ email: imad.hadi@su.edu.krd}

\author{\textbf{Bestoun S. Ahmed*}}
\address{Department of Mathematics and Computer Science, Karlstad University\\ 651 88 Karlstad, Sweden\\ email: bestoun@kau.se}

\author{Moayad Y. Potrus}
\address{Software and Informatics Engineering, Salahaddin University-Erbil (SUE)\\ 44002, Erbil, Iraq
\\ email: moayad.potrus@su.edu.krd}

\author{\textbf{Kamal Z. Zamli}}
\address{Faculty of Computer Systems and Software Engineering Universiti Malaysia Pahang\\ 26300 Gambang, Pahang, Malaysia\\ email: kamalz@ump.edu.my}

\begin{abstract}

To ensure the quality of current highly configurable software systems, intensive testing is needed to test all the configuration combinations and detect all the possible faults. This task becomes more challenging for most modern software systems when constraints are given for the configurations. Here, intensive testing is almost impossible, especially considering the additional computation required to resolve the constraints during the test generation process. In addition, this testing process is exhaustive and time-consuming. Combinatorial interaction strategies can systematically reduce the number of test cases to construct a minimal test suite without affecting the effectiveness of the tests. This paper presents a new efficient search-based strategy to generate constrained interaction test suites to cover all possible combinations. The paper also shows a new application of constrained interaction testing in software fault searches. The proposed strategy initially generates the set of all possible $t-tuple$ combinations; then, it filters out the set by removing the forbidden $t-tuples$ using the base forbidden tuple (BFT) approach. The strategy also utilizes a mixed neighborhood tabu search (TS) to construct optimal or near-optimal constrained test suites. The efficiency of the proposed method is evaluated through a comparison against two well-known state-of-the-art tools. The evaluation consists of three sets of experiments for 35 standard benchmarks. Additionally, the effectiveness and quality of the results are assessed using a real-world case study. Experimental results show that the proposed strategy outperforms one of the competitive strategies, ACTS, for approximately 83\% of the benchmarks and achieves similar results to CASA for 65\% of the benchmarks when the interaction strength is 2. For an interaction strength of 3, the proposed method outperforms other competitive strategies for approximately 60\% and 42\% of the benchmarks. The proposed strategy can also generate constrained interaction test suites for an interaction strength of 4, which is not possible for many strategies. The real-world case study shows that the generated test suites can effectively detect injected faults using mutation testing.

\end{abstract}
\end{frontmatter}

\section{Introduction}

Software testing plays a significant role in specifying the quality of developed software. High-quality software can be achieved by employing an intensive testing process to detect most of the bugs. Ideally, identifying all possible bugs is achievable by applying exhaustive testing with a large test suite. However, exhaustive testing leads to more time spent in software testing, which increases the development cost. Software that is released without being tested well (due to a release deadline constraint or any other reasons) loses trust in the market. As estimated in \cite{KevinDunne2016}, software faults can generate a drop in the product stock price of $4 \: - \: 6\%$ on average (for companies encountering multiple software failures) and further generate almost 3 billion dollars of market losses. Hence, software testers should generate high-quality test suites (with a minimum number of test cases that cover most of the bugs) to develop high-quality software \cite{Ahmed2016}.


Recent developments in the industry have defined the need for the production of configurable software systems to cope with variability in the options and the system environment \cite{YilmazIEEETrans}. These types of software systems consist of many components or modules that interact with each other. Each component has many configurations and options. The components are integrated to produce new software configurations based on the user's needs. Common examples of a configurable system include software product lines (SPLs) (reusing the available component from a set that share the same core to produce new software), web servers or databases, and compilers that can be configured with command-line arguments or configuration files \cite{BestounASOC}. Evidence has shown that software faults or bugs occur during a small number of interactions between the configuration options of these components \cite{Kuhn2013}. This forces a software tester to test combinations of all the configuration options. However, most of the time, testing all the combinations is not always feasible due to the many possible test cases. Instead, a systematic sampling technique is used in the literature to select test cases that cover all the combinations effectively. This technique dramatically reduces the size of the exhaustive test suite in such a way that covers all possible combinations at least once. This testing technique is called combinatorial interaction testing (CIT). CIT has been introduced as an active research area, and many excellent research outcomes have been achieved in the last decade. CIT has also been applied in many industrial case studies \cite{Ahmed2017,BestounQRSS}. In an application of CIT, several studies found that, in reality, most of the time, there are constraints between the input parameters. These constraints lead to the generation of invalid test cases when they are skipped or violated \cite{Lin2016,Ahmed2017a}. In fact, the majority of the research studies in the literature have focused on CIT without considering the constraints.

Recently, the research focus has shifted to the practical applications of CIT. Since most real-world systems have many constraints among the input configurations, constrained CIT (CCIT) can be seen as a current trend and future direction of CIT \cite{Petke20152,Ahmed2017,YilmazIEEETrans}. CCIT is considered a part of CIT that is a nondeterministic polynomial time (NP-hard) problem \cite{Nie2011}. Many approaches have been proposed for CIT in the literature, and a few of them support constraints. There are two main approaches that are used to generate CCIT test suites: greedy algorithms and metaheuristic search algorithms. In both cases, integrated methods are used to resolve the constraints. The greedy and metaheuristic approaches generate solutions iteratively. Most greedy strategies are deterministic, whereas metaheuristic algorithms are nondeterministic. Due to the nature of greedy algorithms, they run faster than metaheuristic strategies. However, metaheuristic strategies generate the test suites efficiently in terms of test suite size. Despite having many approaches for solving CIT, there is no general strategy for generating an optimum test suite, and each approach has advantages and disadvantages. Thus, each approach can be used for a different type of system and configuration \cite{Ahmed2017a}.


Evidence in the literature has shown that most of the constraint-handling CIT strategies in recent works, such as \cite{Lin2016,Yamada2016,Garvin2011}, were based on constraint solvers such as an SAT solver integrated with metaheuristic algorithms. However, the main problem with these solvers is that they do not scale well, especially for highly configurable systems with a large number of constraints. Another issue is that they operate as a black box and are hard to customize. Forbidden-tuple-based strategies, however, are more scalable for constraint handling and have been used with greedy algorithms. Metaheuristic algorithms such as the tabu search (TS) \cite{Gonzalez-Hernandez2010org} or simulated annealing \cite{Cohen2003} have better efficiency in generating a smaller test suite than greedy algorithms. To fill this gap, we combine the efficiency of a metaheuristic algorithm with the scalability of a forbidden-tuple-based constraint-handling strategy.

In this paper, we propose a new metaheuristic method that uses the mixed neighborhood functions of the TS to generate constrained test suites. We also present an implementation of a forbidden-tuple-based approach as a new algorithm for constraint handling. The reasons behind choosing the mixed neighborhood TS as a search algorithm are twofold: first, in the literature, this algorithm has reported new bound results for unconstrained CIT \cite{Gonzalez-Hernandez2010org,Gonzalez-Hernandez2015}; second, the simplicity of the algorithm may lead to less computation. Another reason for selecting forbidden-based constraint handling is the scalability and efficiency that the strategy showed in \cite{Czerwonka2008,Yu2014,Yu2015}. To meet the goal of this paper, we also modify the original mixed TS proposed by \cite{Gonzalez-Hernandez2010org} to improve the neighborhood functions of the TS with random selection techniques instead of sequential selection. Additionally, we use a new alternative for the construction of the initial solution of the constrained mixed covering array (CMCA) to achieve better convergence. To diversify the search space of the TS algorithm, we optimize one of the neighborhood functions from a local best evaluation into the global best assessment. Furthermore, we integrate the forbidden-tuple-based constraint-handling strategy with a mixed neighborhood TS to replace the constraint solver. To verify the quality and effectiveness of the generated test suites by the proposed method, we carry out an empirical case study for mutation testing.

The remaining sections of this paper are structured as follows: Section \ref{Background} starts with an example to introduce CIT and the constrained interaction testing problem; then, we show the related mathematical notions. Section \ref{Method} presents the detailed steps and algorithms of the proposed methods for both the test case generation and constraint handling. Section \ref{Evaluation} presents an experimental evaluation and discusses the evaluation of the efficiency and effectiveness of the proposed method. Section \ref{Threats} highlights the potential challenges to the validity of the experimental results. Section \ref{RelatedWork} lists and summarizes the related works. Finally, Section \ref{Conclusion} concludes this paper.

\section{Background}\label{Background}

This section gives an overview of the theoretical backgrounds on the CCIT approach and how to model the system-under-test (SUT) using this approach. Also, the section explains the general concepts and mathematical notation of constraints interaction testing.

\subsection{Constrained Combinatorial Interaction Testing}

CCIT is a type of CIT in which there are constraints among the input parameters. To demonstrate the concept of CIT and constraint support, suppose there is an SUT that has a set of $k$ input parameters such that $P = \{ p_1,p_2,..,p_k \}$ and each parameter has a set of $n$ values or options such that ${V_{p_i}} = \{ v_1,..,v_{n} \}$ is equivalent to the domain of $p_i$. Often, configurable systems have constraints among their parameter values. As mentioned in \cite{Kuhn2015}, an important step of modeling the input parameters is to identify possible constraints that may occur among different parameters and values in such a way that these constraints can be automatically processed. Every test case must satisfy all the identified constraints; if not, a test may be rejected by the SUT and thus would not serve the purpose. Similar to parameter and value identification, constraints can be elicited from different sources of information, e.g., required documents, feature models, the system environment, and domain knowledge.

Regarding the modeling of the constraints, two popular forms of constraint representations have been used in the literature, including logical expressions (boolean satisfiability formulas) \cite{Cohen2007,Yamada2015,Lin2016} and forbidden tuples \cite{Czerwonka2008,Yu2014,Yu2015}. A common way of specifying constraints is to represent the constraints as a set of forbidden tuples, i.e., combinations that cannot appear in any test. Thus, as mentioned in \cite{Yu2014}, forbidden tuples are equivalent to conjunctive normal form (CNF) constraints. A test is valid if and only if it does not contain a forbidden tuple \cite{Kuhn2015}

Consequently, let us assume that the SUT has $m$ constraints among their parameter values that can be represented as a set of forbidden tuples $C = \{ {c_1,c_2,..,c_m} \}$, where each tuple $c_j \in C$ indicates a set of invalid combinations of parameter values and each $c_j = \{ (p_x,v_x), (p_y,v_y), ... \}$ is a forbidden tuple such that each element in $c_j$ is in the form of a parameter-value pair. For instance, $p_x = v_x$ and $p_y = v_y$ indicate an invalid combination between both parameters $p_x$ and $p_y$ that is not allowed to appear in any test case.

For example, consider the model in Table \ref{ConfigurableSystemExampleTable} as an SUT that illustrates the environmental configuration of Drupal 8 CMS, which has only $4$ configuration parameters. The first three parameters have three values or options, and the last parameter has two values or options. Additionally, three constraints can be identified from the SUT environment. The constraints are among the parameters $OS$, $Browser$ and $Database$, since $macOS$ and $Linux$ do not support the browser $MS \; Edge$, and $macOS$ does not support the database $MS \; SQL$; these constraints are restrictions that cannot appear in any test case. As a result, the input parameters and constraints of the current SUT are modeled as follows:

\begin{align*} 
P &= \{OS,Browser, Database, Server\}\\
V_{os} &= \{Windows, Linux, macOS\}\\
V_{Browser} &= \{Firefox, Chrome, Ms \; Edge\}\\
V_{Database} &= \{MySQL, PostgreSQL, MS \; SQL\}\\
V_{Server} &= \{Apache, Nginx\}\\
C &= \{\{ (OS, Linux), (Browser,MS \; Edge)\},\\
& \qquad \! \{(OS,macOS), (Browser, MS \; Edge)\},\\
& \qquad \! \{(OS,macOS), (Database, MS \; SQL)\}\}
\end{align*} 

\begin{table}
\caption{\label{ConfigurableSystemExampleTable}Environment configuration for Drupal 8 CMS as a SUT
model}
\centering%
\begin{tabular}{|c|c|c|c|c|c|}
\cline{3-6} \cline{4-6} \cline{5-6} \cline{6-6} 
\multicolumn{1}{c}{} &  & \multicolumn{4}{c|}{Configuration Parameter} \tabularnewline
\cline{2-6} \cline{3-6} \cline{4-6} \cline{5-6} \cline{6-6} 
\multicolumn{1}{c|}{} & \# & Operating System (OS) & Browser & Database & Server \tabularnewline
\hline 
\multirow{3}{*}{\rotatebox[origin=c]{270}{Values}} & 0 & Windows & Firefox & MySQL & Apache \tabularnewline
\cline{2-6} \cline{3-6} \cline{4-6} \cline{5-6} \cline{6-6} 
 & 1 & Linux & Chrome & PostgreSQL & Nginx \tabularnewline
\cline{2-6} \cline{3-6} \cline{4-6} \cline{5-6} \cline{6-6} 
 & 2 & macOS & MS Edge & MS SQL & \tabularnewline
\hline 
\multicolumn{6}{|l|}{
\begin{tabular}{l}
Constraints:\tabularnewline
1. Linux system \textbf{does not support} Microsoft Edge browser.\tabularnewline
2. macOS system \textbf{does not support} Microsoft Edge browser.\tabularnewline
3. macOS system \textbf{does not support} Microsoft SQL Server.\tabularnewline
\end{tabular}
} \tabularnewline
\hline 
\end{tabular}
\end{table}

\begin{figure}
\centering\includegraphics[scale=0.5]{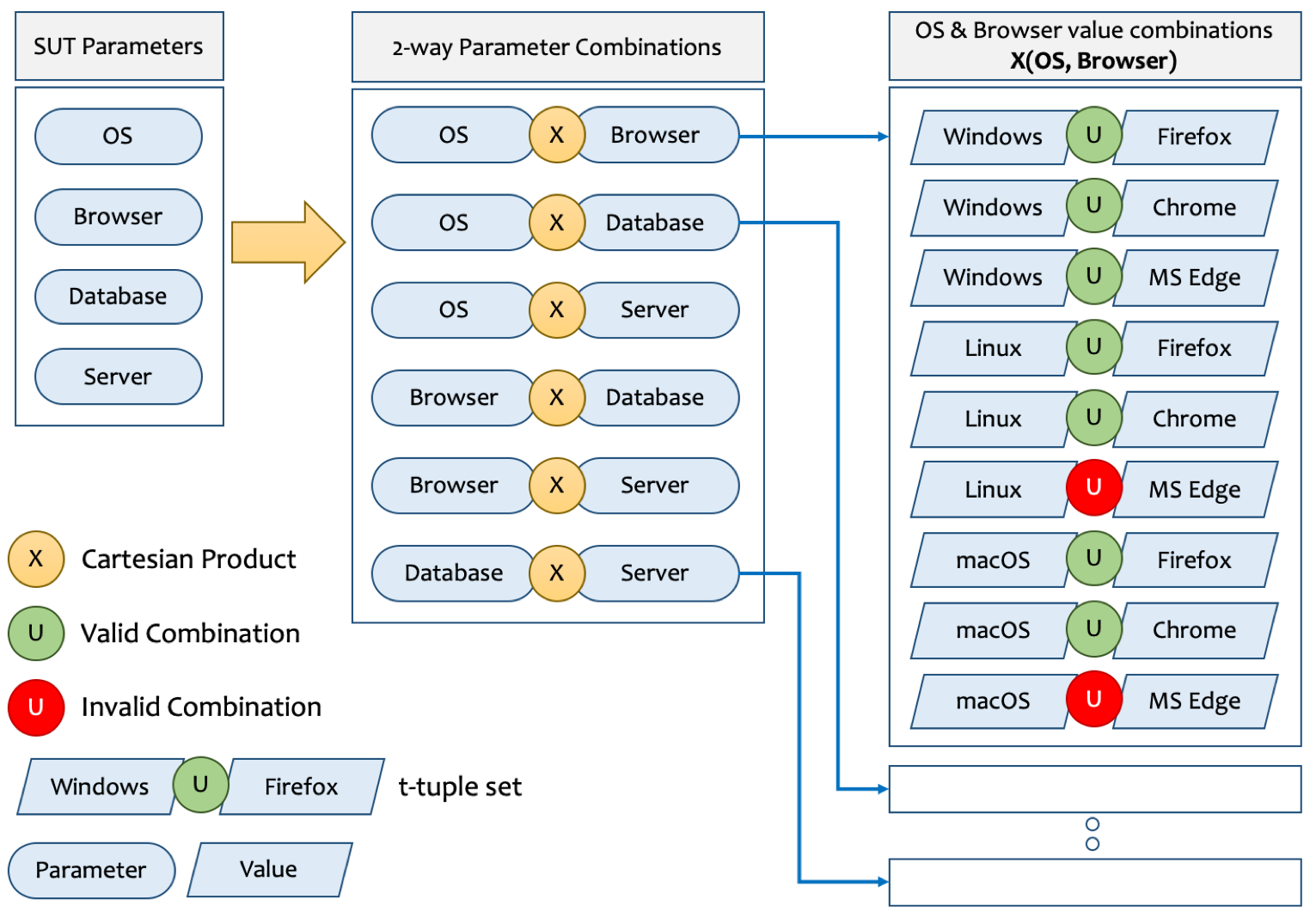}

\caption{\label{ExampleGraph} The pairwise modeling of the SUT in Table \ref{ConfigurableSystemExampleTable}}
\end{figure}

CIT is a combination of parameter values with a specific interaction strength (or combination degree) usually denoted as $t$. The SUT model in Table \ref{ConfigurableSystemExampleTable} has $3^3 \times 2^1 = 54$ possible combinations, where in the first number $3^3$, the base number indicates the first three parameters and the exponent refers to three values for each parameter. The second number $2^1$ indicates that the last parameter has two values. For this small system, it could be feasible to consider all possible test cases. However, when the configuration of the SUT grows, for example, with 12 parameters and each having 4 values, $16,777,216$ test cases are needed to test all possible configurations, which is impossible due to time and budget constraints. Here, CIT techniques will reduce the number of test cases dramatically. For example, in the current SUT, instead of applying all 54 test cases to cover all possible configurations, we only need 10 test cases to cover all possible t-tuple configurations at least once, as shown in Table \ref{TwayConfigurationTable}. To understand the interaction strength $t$, let $X$ be a function that takes $t$ parameter combinations and then computes the Cartesian product of their possible values that produces t-tuple sets. For example, $t=2$ means the combination of two parameters (sometimes called pairwise), such as $X(OS,Browser)$, $X(OS,Database)$, $X(OS,Server)$, $X(Browser,Database)$, $X(Browser,Server)$ and $X(Database,Server)$. Figure \ref{ExampleGraph} shows the pairwise model and the products of combinations for the SUT in Table \ref{ConfigurableSystemExampleTable}.

Table \ref{TwayConfigurationTable} shows a test suite that equivalents to pairwise CIT of the SUT, where each row represents a test case, and each column is an input parameter and contains only values from its domain. This combination structure can be represented with the help of a mathematical object called Covering Array (CA) as defined in the next subsection.

\begin{table}
\caption{\label{TwayConfigurationTable} 2-way CCIT of SUT}

\centering%
\begin{tabular}{|c|c|c|c|c|}
\hline 
Test & OS & Browser & Database & Server\tabularnewline
\hline 
\hline 
1 & Windows & Chrome & MySQL & Apache\tabularnewline
\hline 
2 & Linux & Firefox & PostgreSQL & Apache\tabularnewline
\hline 
3 & Linux & Chrome & MS SQL & Nginx\tabularnewline
\hline 
4 & macOS & Firefox & MySQL & Nginx\tabularnewline
\hline 
5 & Windows & MS Edge & PostgreSQL & Nginx\tabularnewline
\hline 
6 & Windows & MS Edge & MS SQL & Apache\tabularnewline
\hline 
7 & macOS & Chrome & PostgreSQL & Apache\tabularnewline
\hline 
8 & Linux & Chrome & MySQL & Nginx\tabularnewline
\hline 
9 & Windows & Firefox & MS SQL & Nginx\tabularnewline
\hline 
10 & Windows & MS Edge & MySQL & Apache\tabularnewline
\hline 
\end{tabular}
\end{table}


\subsection{Constrained Covering Array (CCA)}

A covering array (CA) is a mathematical object denoted by $CA(N; t, k, v)$ that is an $N$ row by $k$ column (parameter) array. The key feature of a CA is that for every $N \times t$ subarray, all possible value combination $t-tuple$ sets appear at least once and are considered to be covered tuples, where $N$ represents the number of tests, $k$ is the number of parameters, each parameter has $v$ values, and $t$ is the interaction strength.

A mixed CA (MCA) denoted by $MCA(N; t, k, V_{p_1}...V_{p_k})$ is an extended version of a CA. The only difference between an MCA and a CA is that the domain parameters in the MCA are nonunified, since the domain of the parameters in the CA is unified. The values in the $i^{th}$ column belong to the set $V_{p_i}$. For example, according to the MCA definitions, the test suite in Table \ref{TwayConfigurationTable} can be represented as $MCA(10,2,4,3^3 2^1)$.

CA and MCA can be extended to a Constrained Covering Array (CCA) and Constrained Mixed Covering Array (CMCA) by including the $C$ constraints set variable to their definitions that can be written as $CCA(N; t, k, v, C)$ and $CMCA(N; t, k, (V_{p_1}...V_{p_k}), C)$. According to the new definitions, a Boolean variable is added to all $t-tuple$ sets that indicate the validity of the tuple, in a way that invalid tuples are flipped to false if $t-tuples$ is a subset of $C$. Otherwise, valid tuples are true; this variable is used to satisfy the validity of generated rows (test cases) during the test suite generation, as shown in Figure \ref{ExampleGraph}. For example, invalid $t-tuples$ such as $\{(OS,macOS),(Browser, MS \; Edge)\}$ and $\{(OS,Linux),(Browser, MS \; Edge)\}$ are in red color, and the valid tuples are in green color.

\section{The Proposed Method}\label{Method}

This section describes the details of the proposed method that combines a meta-heuristic algorithm Tabu Search (TS) as a search technique with a custom constraint handling technique known as Base Forbidden Tuple (BFT). The upcoming subsections describe the components of the proposed method individually. First, we explain the concept of Tabu Search Algorithm and the properties of it briefly, then, we present the steps of constraint handling technique, after that, we demonstrate the detailed steps of the test case generation.

\begin{algorithm}[H]
\KwIn {$ problem \: X$}
\KwOut{$ optimized \:  X$}

$T \gets \phi$ \tcp*{empty tabu list}

$s \in S$ \tcp*{candidate solution}

$s_{best} \gets s$ \tcp*{assume s is the best solution}

$T \gets s$

\While{$f(s_{best}) > 0$}{

	$s' \gets N(s,T)$
	
	\If{ $f(s') \leq f(s_{best})$ }{
	
		$s_{best} \gets s'$
		
	}
	
	$T \gets s'$
	
	$s \gets s'$

}
\caption{\label{TSBaseicAlgo} Tabu Search Algorithm}
\end{algorithm}

\subsection{The Tabu Search}

The tabu search (TS) algorithm is a metaheuristic local search algorithm initially proposed by Glover in 1986. \cite{Glover1986}. The central concept of the TS is to integrate an adaptive memory into a search space called a tabu list \cite{Glover1997}. The tabu list is a queue data structure that stores a specific number of recent movements that are performed to change a current solution $s$ to a new solution $s'$. When a new solution is generated, the TS algorithm avoids the movements that exist in the tabu list. The current movement is queued to the tabu list, and the oldest movement is dequeued from the list so that the current movement is forbidden for the next generation for a specific number of iterations, which is based on the queue size. Sometimes, the TS allows a movement that is in the tabu list, which can produce better solutions than the current best; this is another characteristic of the TS algorithm called aspiration criteria. Other essential components of the TS are the intensification and diversification search strategies. Intensification is a strategy that intensively searches regions where the best solutions have been found to find better solutions. In contrast, the diversification strategy visits the unexplored regions of the search space with the hope of finding new solutions that may vary in different ways from the previous solutions.

Algorithm \ref{TSBaseicAlgo} shows the pseudocode of the basic TS procedure. The algorithm presents the main components of the TS, including the 1) initialization of a subset of the search space $s$; 2) tabu list queue $T$; 3) neighborhood function $N(s,T)$ that transforms the current solution $s$ to another new solution $s'$; 4) evaluation function $f(s)$; and 5) stopping criteria $f(s_{best}) > 0$, which in this case is the minimization function. Note that the keyword movement and transformation are used interchangeably in the upcoming subsections.

\subsection{Constraint handling}

As mentioned earlier, configurable systems may have different constraints within a combination of configuration options. Without handling these constraints, the generated test cases may contain invalid combinations, which leads to a low-quality and incomplete test suite \cite{Ahmed2015,Ahmed2017a}. Before providing the details of the proposed constraint-handling algorithm, it is important to understand the impact of constraints on the CIT sampling sizes and the complexity of the problem.

As mentioned in \cite{Cohen2007}, constraints minimize the number of feasible system configurations, but they might increase or decrease the sample size. The impact on the sample size depends on the interplay between the number and type of constraints and the model of the SUT. For detailed information on the impact of constraints, we refer the reader to \cite{Cohen2007}.

\subsubsection{The base forbidden tuple algorithm}

Usually, it is impossible to directly use the input set of forbidden tuples to check the validity of the test cases. This is because user-input forbidden tuples may imply more hidden forbidden tuples that are not explicitly specified \cite{Cohen2008}. It is not feasible for users to identify all forbidden tuples in an SUT. However, if the testers can generate all the hidden forbidden tuples, the validity of the test cases can be easily determined \cite{Yu2015}.

As pointed out in \cite{Cohen2007}, the number of hidden forbidden tuples might be very large and impossible to identify by the users. To overcome this problem, in this work, we propose a new algorithm for constraint handling called the base forbidden tuple (BFT) algorithm; the BFT algorithm is a custom constraint-handling algorithm that is based on two essential operations: (1) a derivation that derives new hidden or implicit forbidden tuples from the initially forbidden tuples and (2) a simplification that removes redundant forbidden tuples after each derivation. This algorithm receives the initial constraints $C$ as an input set of invalid combinations that are a set of forbidden tuples and exports the processed constraints as a set of BFTs. Before explaining the details of these operations, it is necessary to define the forbidden tuple and BFT and the derivation and simplification steps.

\begin{definition} 
\textbf{Tuple:} A Tuple is a set of elements, each element is in the form of parameter value pair $(p,v)$.
\end{definition}

A tuple of size $r$ represents the value assignment of $r$ parameter-value pairs. But, it is important to note that a tuple can only contain one value from the same parameter. Tuples can be either valid or invalid based on the identified constraints from SUT model.

\begin{definition}
\textbf{t-tuple:} A t-tuple is a tuple that consists of $t$ elements.
\end{definition}

A t-tuple is a set of $t$ elements, where $t$ refers to an interaction strength, and each element is represented as a parameter-value pair. Additionally, the same parameter can only have one value in the set.

\begin{definition}
\textbf{Valid Tuple:} A Valid Tuple is a tuple that does not belong to the set of constraints $C$ and contains no BFT.
\end{definition}

Any tuple that does not belong to the processed set of constraints $C$ and it is not a forbidden tuple allowed to be appeared in the test cases. 

\begin{definition}
\textbf{Forbidden Tuple:} A Forbidden Tuple is any tuple that belongs to the set of constraints $C$ which is an invalid combination of parameter options.
\label{def4}
\end{definition}

Based on Definition \ref{def4}, a forbidden tuple is an invalid combination of configurations or system features that cannot exist in any test case. Another observation is that forbidden tuples can be further processed and simplified to extract hidden tuples from them until they converge into a set of BFTs.

\begin{definition}
\textbf{Base Forbidden Tuple:} A Base Forbidden Tuple (BFT) is a forbidden tuple that is unique and can not contain any other forbidden tuple.
\label{def5}
\end{definition}

Based on Definitions \ref{def4} and \ref{def5}, it can be observed that every BFT is a forbidden tuple. Accordingly, it is obvious that a given forbidden tuple contains at least one BFT. Additionally, any tuple that contains a forbidden tuple should be a forbidden tuple. Sometimes even if a tuple does not contain any forbidden tuple, it may still be a forbidden tuple. For example, consider the following SUT that receives three binary parameters and two forbidden tuples as initial set of constraints:

\begin{align*} 
P &= \{p_1,p_2, p_3\}\\
V_{p_1} &= \{0, 1\}\\
V_{p_2} &= \{0, 1\}\\
V_{p_3} &= \{0, 1\}\\
C &=\{\{(p_1,1),(p_2,0)\}, \{(p_2,1),(p_3,1)\} \}
\end{align*} 


Here, the input constraints contain two forbidden tuples $\{(P_1,1),(P_2,0)\}$ and $\{(P_2,1),(P_3,1)\}$, but still this tuple $\{(P_1,1),(P_3,1)\}$ can be considered as a forbidden tuple. However, it contains none of the forbidden tuples in $C$. The reason behind this situation is that, for any test case that contains this tuple $\{(P_1,1),(P_3,1)\}$, if $P_2 = 0$ violates the first forbidden tuple $\{(P_1,1),(P_2,0)\}$ and if $P_2 = 1$, then it will violate the second forbidden tuple $\{(P_2,1),(P_3,1)\}$. Hence, the input forbidden tuples $C$ should be processed and converted into a set of BFTs such as $C =\{ \{(P_1,1),(P_2,0)\}, \{(P_2,1),(P_3,1)\}, \{(P_1,1),(P_3,1)\} \}$. According to the above example, it is clear that the input set of forbidden tuples cannot directly be used to check the validity, and every forbidden tuple in the initial $C$ should be converted into a set of BFTs.

It is important to note that the BFT generation process from an initial set of forbidden tuples $C$ in the proposed algorithm depends on two primary iterative operations: the derivation of the hidden constraints and the simplification process after each derivation process. Each operation has a specific definition that is a core function of the proposed constraint-handling algorithm, as discussed below.

\begin{definition}{\textbf{Hidden forbidden tuple derivation:}} Given two forbidden tuples $X$ and $Y$ that $(p_i,x) \in X$ and $(p_i,x') \in Y$ , $Z$ can be derived as a new forbidden tuple as follows $Z = X \cup Y \setminus \{(p_i,V_{p_i})\}$ which is a hidden forbidden tuple.
\label{def6}
\end{definition}

\begin{table}[]
    \small
    \centering
    \caption{ \label{bft-derive} Possible Scenarios of deriving new forbidden tuples}
    \begin{tabular}{|l m{3.5cm} >{\centering\arraybackslash}m{2.5cm}l|}
\hline 
\multicolumn{4}{|c|}{
\begin{tabular}{@{}l@{}}
Parameter: $P = \{P1, P2, P3, P4 \}$ \tabularnewline
Parameter Values: $V_{P1} = \{0,1,2\}, V_{P2} = \{0,1,2\}, V_{P3} = \{0,1,2\}, V_{P4} = \{0,1,2\}$ 
\end{tabular} 
}\tabularnewline
\hline 
1 & %
\begin{tabular}{@{}l@{}}
$\{(\textbf{P1},0), (P3,0)\}$ \tabularnewline
$\{(\textbf{P1},1), (P2,1), (P3,0)\}$ \tabularnewline
$\{(\textbf{P1},2), (P4,2)\}$ \tabularnewline
\end{tabular} & w.r.t. P1 & \{(P2,1), (P3,0), (P4,2)\}\tabularnewline
\hline 
2 & %
\begin{tabular}{@{}l@{}}
$\{(P1,0), (\textbf{P2},0)\}$\tabularnewline
$\{(P1,1), (\textbf{P2},1), (P3,0)\}$ \tabularnewline
$\{(\textbf{P2},2), (P4,0)\}$ \tabularnewline
\end{tabular} & w.r.t. P2 & %

\begin{tabular}{@{}l@{}}
\small{\st{\{(\textbf{P1},0), (\textbf{P1},1), (P3,0), (P4,0)\}}} \tabularnewline
invalid derived tuple \tabularnewline
\end{tabular}  \tabularnewline
\hline 
3 & %
\begin{tabular}{@{}l@{}}
$\{(\textbf{P3},0)\} $\tabularnewline
$\{(\textbf{P3},1)\}$\tabularnewline
$\{(\textbf{P3},2)\}$\tabularnewline
\end{tabular} & w.r.t. P3 & \{ \}\tabularnewline

\hline 
4 & %
\begin{tabular}{@{}l@{}}
\{(\textbf{P1}, \textit{0}), (P2, 0)\}\tabularnewline
\{(\textbf{P1}, \textit{0}), (P2, 2)\}\tabularnewline
\{(\textbf{P1}, \textbf{1}), (P2, 2)\}\tabularnewline
\{(\textbf{P1}, \textbf{1}), (P3, 0)\}\tabularnewline
\{(\textbf{P1}, 2), (P4, 0)\}\tabularnewline
\end{tabular} $\implies $ 
& %
\begin{tabular}{@{}c@{}}
\{(P2, 0), (P2, 2)\}\tabularnewline
 $\times$  \tabularnewline
\{(P2, 2), (P3, 0)\}\tabularnewline
$\times$ \tabularnewline
\{(P4, 0)\}\tabularnewline
\end{tabular} 
& $\implies$ %
\begin{tabular}{@{}l@{}}
\st{\{(\textbf{P2}, 0), (\textbf{P2}, 2), (P4, 0)\}}\tabularnewline
\{(P2, 0), (P3, 2), (P3, 0)\}\tabularnewline
\{(P2, 2), (P4, 0)\}\tabularnewline
\textbf{\{(P2, 2), (P3, 0), (P4, 0)\} } \tabularnewline
\end{tabular}\tabularnewline
\hline 

\end{tabular}
\end{table}

From Definition \ref{def6}, it can be observed that if there are two forbidden tuples that share pairs with the same parameter $p_i$ but different values, then the remaining pair elements in both forbidden tuples $X$ and $Y$ (except the pairs that share the same parameter) can be derived as a hidden forbidden tuple. This definition can be generalized for $m$ tuples that share pairs with the same parameter and different values. It is important to note that during the derivation process, there may be a derived tuple that contains different values for the same parameter pairs. In this case, the derived set of pairs is an invalid tuple, and no forbidden tuple is derived from this invalid tuple.

For example, consider the simple SUT in Table \ref{bft-derive} with three different scenarios for deriving new forbidden tuples adopted from \cite{Yu2015}. In the first scenario, a new forbidden tuple is derived from three forbidden tuples with respect to the parameter $P1$. In the second scenario, the derivation is with respect to $P2$. As a result, there is a conflict between the two pairs $(P1,0)$ and $(P1,1)$ that share $P1$ in the newly derived set of pairs. Thus, it's an invalid tuple. The third scenario is a specific case in which an empty tuple is generated from 3 forbidden tuples of size 1. It actually refers to the fact that the parameter $P3$ cannot be assigned with these valid values.

Another important derivation rule that should be highlighted is that when the same parameter-value pair appears in different forbidden tuples in $C$, there will be a possibility to derive multiple new forbidden tuples with respect to the shared parameter. For example, in the last scenario from Table \ref{bft-derive}, there are five forbidden tuples, and the parameter-value pair $(P1,0)$ appears in the first and second forbidden tuples. The parameter-value $(P1,1)$ pair appears in the third and fourth forbidden tuples. However, the pair $(P1,2)$ appears only in the last tuple. The remaining pairs of forbidden tuples are grouped based on shared pairs; for example, $\{(P2, 0), (P2, 2)\}$ is grouped based on $(P1,0)$, $\{(P2, 2), (P3, 0)\}$ is grouped according to $(P1,1)$, and the set $\{(P4, 0)\}$ is grouped individually based on $(P1,2)$. Then, the cross product of these groups is calculated. As a result, multiple forbidden tuples are derived. Note that the first derived set of pairs is not a valid tuple as it contains two pairs of the same parameter and will be removed. The three remaining sets of pairs are newly derived forbidden tuples.

\begin{definition}{\textbf{Simplification:}} Given tuple $T$ can be removed from a set of forbidden tuples $C$, if there exist another tuple $T'$ such that $T' \subseteq T$.
\label{def7}
\end{definition}

The simplification process is an iterative process that removes any forbidden tuple that is repeated or is a part of another forbidden tuple in $C$. For example, in the last scenario from Table \ref{bft-derive}, the last forbidden tuples $\{(P2,2), (P3,0), (P4,0)\}$ in bold cannot be a BFT as it covers the third forbidden tuple $\{(P2, 2), (P4, 0)\}$. The result of the validity check will not be affected if we remove the last forbidden tuple because any test case that satisfies that forbidden tuple must also satisfy the third forbidden tuple.

\begin{table}[]
    \centering
    \caption{\label{bft-derive-example} Example of BFT generation process}
    {\footnotesize{}}%
\begin{tabular}{|c|lcr|}
\hline 
\multicolumn{4}{|c|}{{\footnotesize{}}%
\begin{tabular}{l}
{\footnotesize{}Parameter: $P = \{ P1, P2, P3, P4 \}$} \tabularnewline
{\footnotesize{}Parameter Values: $V_{P1}=\{0,1,2\}, V_{P2}=\{0,1,2\}, V_{P3}=\{0,1,2\}, V_{P4}=\{0,1,2\}$} \tabularnewline
\end{tabular}} \tabularnewline
\hline 
{\footnotesize{}Step 1} & {\footnotesize{}}%
\begin{tabular}{@{}l@{}}
{\footnotesize{}\{(}\textbf{\footnotesize{}P1}{\footnotesize{}, 0),
(P2, 0)\}}\tabularnewline
{\footnotesize{}\{(}\textbf{\footnotesize{}P1}{\footnotesize{}, 0),
(P2, 2)\}}\tabularnewline
{\footnotesize{}\{(}\textbf{\footnotesize{}P1}{\footnotesize{}, 1),
(P4, 0)\}}\tabularnewline
{\footnotesize{}\{(}\textbf{\footnotesize{}P1}{\footnotesize{}, 2),
(P4, 0)\} }\tabularnewline
{\footnotesize{}\{(P2, 1), (P4, 0)\}}\tabularnewline
\end{tabular} & {\footnotesize{}derive w.r.t P1} & {\footnotesize{}}%
\begin{tabular}{@{}c@{}}
{\footnotesize{}\{(P2, 0), (P4, 0)\}}\tabularnewline
{\footnotesize{}\{(P2, 2), (P4, 0)\}}\tabularnewline
\end{tabular}\tabularnewline
\hline 
{\footnotesize{}Step 2} & {\footnotesize{}}%
\begin{tabular}{@{}l@{}}
{\footnotesize{}\{(P1, 0), (}\textbf{\footnotesize{}P2}{\footnotesize{},
0)\}}\tabularnewline
{\footnotesize{}\{(P1, 0), (}\textbf{\footnotesize{}P2}{\footnotesize{},
2)\}}\tabularnewline
{\footnotesize{}\{(P1, 1), (P4, 0)\}}\tabularnewline
{\footnotesize{}\{(P1, 2), (P4, 0)\}}\tabularnewline
{\footnotesize{}\{(}\textbf{\footnotesize{}P2}{\footnotesize{}, 1),
(P4, 0)\}}\tabularnewline
\end{tabular} & {\footnotesize{}derive w.r.t P2} & {\footnotesize{}\{(P1, 0), (P4, 0)\}}\tabularnewline
\hline 
{\footnotesize{}Step 3} & \multicolumn{3}{c|}{{\footnotesize{}Simplification is done with out removing any tuple}}\tabularnewline
\hline 
{\footnotesize{}Step 4} & {\footnotesize{}}%
\begin{tabular}{@{}l@{}}
{\footnotesize{}\{(P1, 0), (P2, 0)\}}\tabularnewline
{\footnotesize{}\{(P1, 0), (P2, 2)\}}\tabularnewline
{\footnotesize{}\{(}\textbf{\footnotesize{}P1}{\footnotesize{}, 1),
(P4, 0)\}}\tabularnewline
{\footnotesize{}\{(}\textbf{\footnotesize{}P1}{\footnotesize{}, 2),
(P4, 0)\}}\tabularnewline
{\footnotesize{}\{(P2, 1), (P4, 0)\}}\tabularnewline
{\footnotesize{}\{(P2, 0), (P4, 0)\}{ +}}\tabularnewline
{\footnotesize{}\{(P2, 2), (P4, 0)\}{ +}}\tabularnewline
{\footnotesize{}\{(}\textbf{\footnotesize{}P1}{\footnotesize{}, 0), (P4, 0)\}{ +}}\tabularnewline
\end{tabular} & {\footnotesize{}derive w.r.t. P1} & {\footnotesize{}\{(P4, 0)\}}\tabularnewline
\hline 
{\footnotesize{}Step 5} & {\footnotesize{}}%
\begin{tabular}{@{}l@{}}
{\footnotesize{}\st{\{(P1, 1), (\textbf{P4}, 0)\}}}\tabularnewline
{\footnotesize{}\st{\{(P1, 2), (\textbf{P4}, 0)\}}}\tabularnewline
{\footnotesize{}\st{\{(P2, 1), (\textbf{P4}, 0)\}}}\tabularnewline
{\footnotesize{}\st{\{(P2, 0), (\textbf{P4}, 0)\}}}\tabularnewline
{\footnotesize{}\st{\{(P2, 2), (\textbf{P4}, 0)\}}}\tabularnewline
{\footnotesize{}\st{\{(P1, 0), (\textbf{P4}, 0)\}}}\tabularnewline
\end{tabular} & \multicolumn{2}{l|}{{\footnotesize{}simplify by P4}}\tabularnewline
\hline 
{\footnotesize{}Step 6} & {\footnotesize{}}%
\begin{tabular}{@{}l@{}}
{\footnotesize{}\{(P1, 0), (P2, 0)\}}\tabularnewline
{\footnotesize{}\{(P1, 0), (P2, 2)\}}\tabularnewline
{\footnotesize{}\{(P4, 0)\}}\tabularnewline
\end{tabular} & \multicolumn{2}{c|}{{\footnotesize{}Algorithm stopped and three BFT are generated}}\tabularnewline
\hline 
\end{tabular}{\footnotesize\par}
\end{table}

\subsubsection{BFT generation process}

In this section, the detailed steps of the BFT generation process is shown. The two core operations of deriving hidden forbidden tuples and simplification process are illustrated by a simple example.

Algorithm \ref{BaseForbiddenAlgo} iteratively derives all BFTs based on the received user-specified constraints or initial set of forbidden tuples $C$. By Definition 6, the algorithm first tries to find all the hidden forbidden tuples. After each derivation, the algorithm simplifies the set of forbidden tuples $C$ and checks them using Definition \ref{def7}; then, it removes the redundant forbidden tuples. The algorithm repeats this process until no new tuples are derived, no further simplification is performed and the size of $C$ will remain the same. At this stage, the set of BFTs is ready to be passed to the proposed test generation framework.

To illustrate the steps in Algorithm \ref{BaseForbiddenAlgo}, we use the SUT model shown in Table \ref{bft-derive-example} that is taken from \cite{Yu2015}. The SUT model has four parameters, each with three values and an initial set of constraints $C$ that contains five forbidden tuples. In step 1, the algorithm derives new hidden tuples using $P1$ based on Definition \ref{def6} that produced two new forbidden tuples. Then, the algorithm tries to derive new forbidden tuples using other parameters, as shown in step 2, that derive new tuples using $P2$, where an additional hidden forbidden tuple is derived. Note that new forbidden tuples cannot be derived using both $P3$ and $P4$, as $P3$ is not available and $P4$ can only derive an invalid tuple; see lines 4-8 in Algorithm \ref{BaseForbiddenAlgo}. Then, the newly derived forbidden tuples in both steps are added to $C$; see line 9 in Algorithm \ref{BaseForbiddenAlgo}. After this derivation round, in step 3, the algorithm tries to simplify $C$, and no forbidden tuples are removed as the currently forbidden tuples in $C$ do not satisfy Definition \ref{def7}; see lines 10-12 in Algorithm \ref{BaseForbiddenAlgo}. The next round of derivation starts with an updated $C$, and the newly derived forbidden tuples are marked with a '+' sign. Step 4 derives the new forbidden tuple $\{(P4, 0)\}$ based on the parameter $P1$. In step 5, there are six forbidden tuples in $C$ that share the same parameter-value pair $(P4, 0)$ as the derived new tuples using this pair will produce an invalid tuple; thus, these tuples are removed. In the final step, no further hidden tuples can be derived from the remaining tuples. Hence, all the remaining tuples are BFTs that are passed to the test generation algorithm for a test validity check.

According to the above example, it can be recognized that there is a relationship between the complexity of the initial set of constraints $C$ and the generated set of BFTs. As discussed by Yu et al. in \cite{Yu2015}, the number of forbidden tuples in the initial set of constraints will affect the generation time. When the number of forbidden tuples is small, the generation of the BFT set is very fast, but a large number of forbidden tuples may lead to a time-consuming generation process. Another observation is that the complexity in the relationships among forbidden tuples in the initial set of constraints will also affect the generation time of the BFT set. Because of the complex relations among forbidden tuples in the initial set of constraints, the generation of the BFT set will probably be time-consuming.

\setcounter{AlgoLine}{0}

\begin{algorithm}
\KwIn {initial $C$}
\KwOut{set of BFT $C$}

\Repeat{$size = size\_of(C)$}{

	$size \gets size\_of(C)$
	
	$hidden \gets \phi$
	
	\For{$X \in C$}{
	
		\For{$Y \in C$}{
		
			\If{$pair \in X \: and \: Y \: by \: Definition \: 6$ }{
			
				$Z \gets X \cup Y \setminus \{pair\}$
				
				$hidden \gets hidden \cup \{Z\}$
				
			}
		}
	}
	
	$C \gets C \cup hidden$
	
	\For{$forbidden\_tuple \in C$}{
	
		\If{ forbidden\_tuple satisfy Definition 7 }{
		
			$C \gets C \setminus tuple$
			
		}
	}
}

\caption{\label{BaseForbiddenAlgo} Base Forbidden Tuple generation}
\end{algorithm}

\subsection{Test case generation algorithm}

\begin{figure}[t]
\centering\includegraphics[scale=0.39]{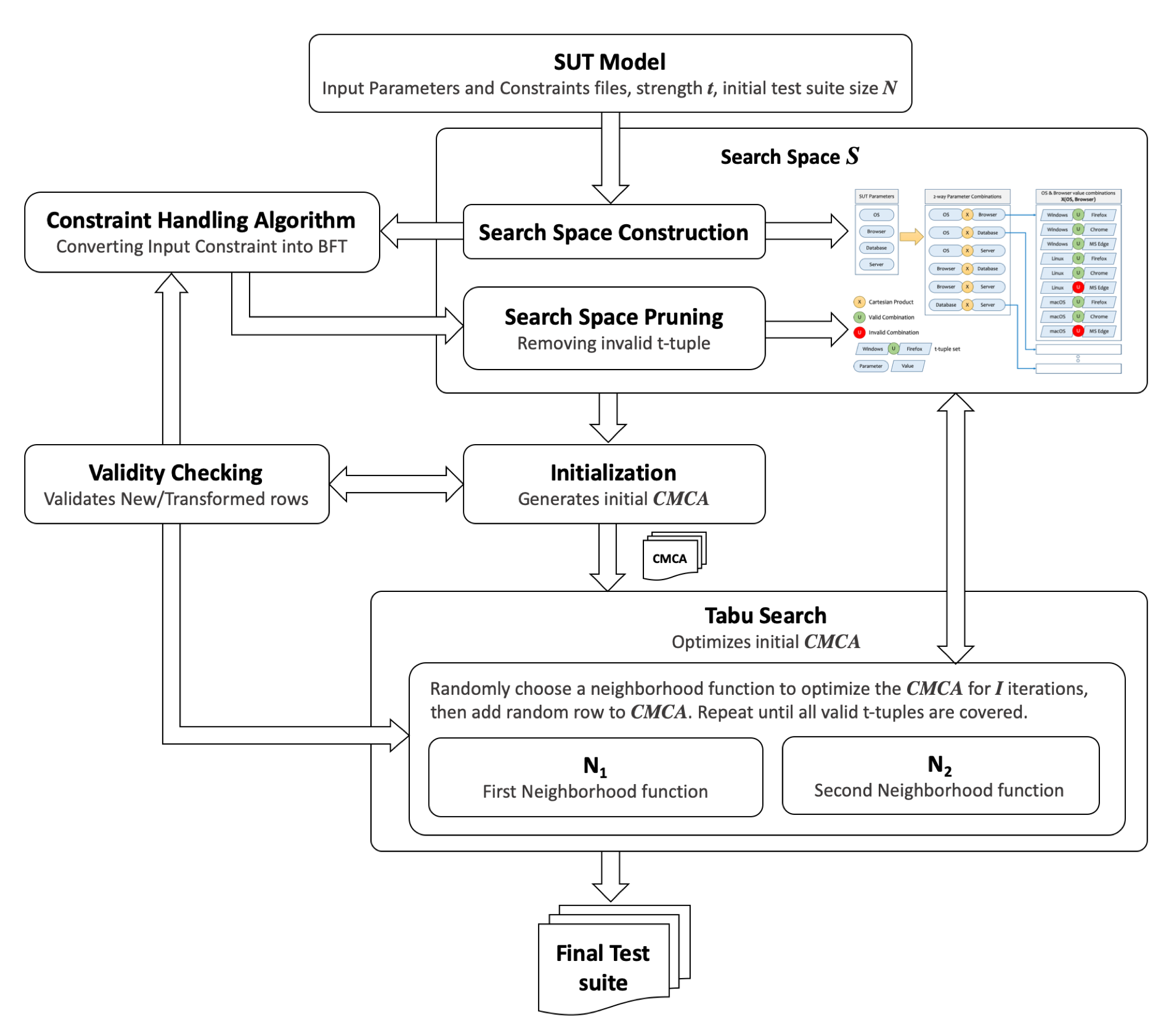}
\caption{\label{BlockDiargam} The overview block diagram for the proposed method}
\end{figure}

Within our strategy, the test case generation algorithm has a special procedure to generate the constrained interaction test suite. As illustrated in Figure \ref{BlockDiargam} first it receives the SUT model files, the coverage strength $t$ and initial size of the test suite $N$. The SUT model file contains the input parameters $P$, values $V_p$, and constraints $C$. Then, the algorithm constructs a search space $S$, which is a two-level hash-table data structure. The first level is the parameter combination and the second level is the combination or cross-product of their values in the form of $t-tuple$ set. Each $t-tuple$ has a flag of a Boolean variable that represents the validity of it. After that, the constraint handling process will start to produce a set of BFT $C$ from the initially forbidden tuples, as shown in Algorithm \ref{BaseForbiddenAlgo}. The algorithm marks every $t-tuple$ in $S$ as a forbidden tuple that is occurred in a set of BFT $C$ as a shown in Algorithm \ref{TestGenerationAlgo} (lines 2 to 5).

\setcounter{AlgoLine}{0}

\begin{algorithm}
\KwIn {SUT($P$, $V_p$, $C$), strength $t$, initial CMCA size $N$, iteration $I$}
\KwOut{complete CMCA $M$}

$S \gets search\_space\_contruction(SUT(P, V_p, C),t)$

\For{$t\_tuple \in S$}{

	\For{$BFT \in C$}{
	
		\If{t\_tuple matches BFT}{
		
			$S[t\_tuple] \gets false$
			
		}
	}
}

$M \gets initialization(N)$\;

\While{ not all valid $t-tuples$ in S are covored }{

	\For{$i\gets I$ \KwTo $0$}{
	
		$M \gets Tabu\_Search(M,I)$
		
		\If{ all valid t-tuples in S are covored }{
		
			$stop$
			
		}
	}
	
	\If{ M is not optimized for some iterations }{
	
		$M \gets new\_row\_with\_hamming\_distance(M)$
		
	}
	
}

\caption{\label{TestGenerationAlgo}Overview of the proposed Algorithm}
\end{algorithm}

Initially, the test case generation process starts with the initialization of the partial CMCA $M$ matrix of size $N$ (line 6) in Algorithm \ref{TestGenerationAlgo}, which describes the initialization process in detail. The TS will start to optimize the initial $M$ for $I$ iterations. After each TS call, the algorithm checks for the $t-tuple$ coverage. If all $t-tuples$ are covered, then the algorithm will stop. Otherwise, it will add a new randomly generated row to $M$. The new row is generated based on a Hamming distance to diversify the search (line 13; see section \ref{InitializationSection} for details). This process is repeated until the strategy covers all the valid $t-tuples$ in $S$. Then, the algorithm will stop and produce a complete CMCA (lines 7-13). In this algorithm, the TS uses two neighborhood functions to explore the search space. Section \ref{NeighborhoodSection} provides details on the neighborhood function and how the algorithm optimizes the initial matrix $M$.


\subsubsection{Initialization}\label{InitializationSection}

The initialization function takes an initial test suite size $N$ and generates a partial test suite CMCA $M$ matrix. The initial test generation has two modes: (1) random generation and (2) a random selection of uncovered valid $t-tuples$. While the first mode involves entirely random generation, the second mode first assigns an empty row, after which the selection method searches for uncovered $t-tuples$ from the search space; then, the found $t-tuples$ are substituted into the empty row, until the row is filled up, and then the row is added to the CMCA $M$ matrix. The combination of the random generation mode with the selection mode helps the TS converge faster.

During random generation mode, the first row of the matrix $M$ is randomly generated such that a value for each parameter $p_i$ is randomly chosen from the set $V_{p_i}$. After this step, the algorithm will check for the validity of this newly generated row. During the validity check, the new row is split into $t-tuple$ combinations. Then, the validity of each $t-tuple$ is checked by looking up the attached boolean variable in the combinatorial data structure. As shown in Figure \ref{ExampleGraph}, if the boolean variable is true, the $t-tuple$ is valid. The $t-tuple$ is invalid if all $t-tuples$ are true. Here, the new row is added to the matrix $M$. An invalid new row means that it contains forbidden tuples or violates the constraints in $C$. The algorithm iterates until the new row is valid, rejecting invalid new rows. From the second row, the algorithm will generate two new random rows known as candidate rows using the same method as for the first row. The difference here is that the algorithm relies on a Hamming distance to choose one of the candidate rows. As demonstrated in Table \ref{Table 3}, the Hamming distance can be computed using Equation \ref{Equation 1}, which is a summation of several different symbols between the candidate row and all rows in the matrix $M$. Hence, the algorithm computes the Hamming distance for the two candidate rows; then, it chooses a candidate row that maximizes the Hamming distance and adds it to the matrix $M$. The aim here is to diversify the values in the new row from the current rows in the matrix to cover more $t-tuples$ and explore the unvisited regions in the search space. The algorithm repeats this process for a specific number of iterations based on the number of covered and uncovered $t-tuples$. Algorithm \ref{Algorithm 4} shows the pseudocode of this algorithm in detail. The random generation mode is described in lines 7-13, whereas the rest of the code refers to the second mode.


\begin{table}
\caption{\label{Table 3} Computing Hamming Distance for candidate rows}

\centering%
\begin{tabular}{|c|c|c|c|l}
\cline{1-4} \cline{2-4} \cline{3-4} \cline{4-4} 
Test & P1 & P2 & P3 & \tabularnewline
\cline{1-4} \cline{2-4} \cline{3-4} \cline{4-4} 
R1 & 0 & 1 & 0 & \tabularnewline
\cline{1-4} \cline{2-4} \cline{3-4} \cline{4-4} 
C1 & 1 & 1 & 0 & $hd(C1) = 1$\tabularnewline
\cline{1-4} \cline{2-4} \cline{3-4} \cline{4-4} 
C2 & 0 & 0 & 1 & $hd(C1) = 2$\tabularnewline
\cline{1-4} \cline{2-4} \cline{3-4} \cline{4-4} 
\multicolumn{1}{c}{} & \multicolumn{1}{c}{} & \multicolumn{1}{c}{} & \multicolumn{1}{c}{} & \tabularnewline
\cline{1-4} \cline{2-4} \cline{3-4} \cline{4-4} 
R2 & 0 & 0 & 1 & $C2$ is selected as a next row\tabularnewline
\cline{1-4} \cline{2-4} \cline{3-4} \cline{4-4} 
\end{tabular}
\end{table}

\begin{align}
hd(r_{s}) & =\mathrel{{\displaystyle \sum_{i=1}^{s-1}\sum_{j=1}^{k}g(m_{i,j},m_{s,j})}}\label{Equation 1}\\
where & \quad g(m_{i,j},m_{s,j})=\begin{cases}
1 & m_{i,j}\not=m_{s,j}\\
0 & otherwise
\end{cases}\nonumber 
\end{align}

\setcounter{AlgoLine}{0}

\begin{algorithm}
\KwIn {N}
\KwOut{partial CMCA M}

let $CTuples()$ = No. of covered $t-tuples$ in $S$

let $UTuples()$ = No. of uncovered $t-tuples$ in $S$

let $HD(R)$ = hamming distance from $R$ to $M$

let $random\_row()$ = creates a valid random row

$M \gets random\_row()$

\For{$i\gets 0$ \KwTo $N$}{

	\tcp{random mode}
	
	\eIf{$UTuples() < CTuples()$}{
	
		$C1 \gets random\_row()$
		
		$C2 \gets random\_row()$
		
		\eIf{$HD(C1) > HD(C2)$}{
		
			$M \gets C1$
			
		}{
		
			$M \gets C2$
			
		}
	}{
	
		\tcp{selection mode}
		
		$temp\_row \gets \{-1\} * size(P)$
		
		let $param\_combinations$ = \{set of paramter combinations\}
		
		\While{temp\_row contatins -1}{
		
	 	   \ForEach{ $combination \in param\_combinations$}{
	 	   
				$t\_tuple \gets select\_t\_tuple(combination)$
				
				$temp\_row[combination] \gets t\_tuple$
				
	        }
		}
		
		$M \gets temp\_row$
	}
}

\caption{\label{Algorithm 4} Initialization function}
\end{algorithm}


As illustrated in Algorithm \ref{Algorithm 4}, during the random selection of the uncovered $t-tuples$, first, the algorithm defines a row of "-1"s, and then the selection mode tries to replace the "-1"s from the row with the selected uncovered valid $t-tuples$ from the search space $S$. Each time a $t-tuple$ is selected, it will be replaced with $t$ "-1"s in the new row until no "-1"s remain. In this mode, there is no need to check the validity for the new row since it is constructed incrementally from the only selected valid $t-tuples$ in the search space. For the remaining iterations, this process is repeated until the size of the matrix $M$ reaches $N$ rows.

\subsubsection{Neighborhood functions}\label{NeighborhoodSection}

In our proposed strategy, the TS algorithm uses two neighborhood functions $N_1$ and $N_2$ to explore the search space and to move from one state to another state as each neighborhood function has a specific move strategy. It is worth to mention that we used two of the same neighborhood functions that originally proposed by \cite{Gonzalez-Hernandez2010}, we also modified the functions to improve the performance (run-time) using a random selection techniques that randomly select a specific number of rows from CMCA $M$ and transform them, instead of transforming all the rows with sequential selection, this improvement helped the neighborhood functions to explore unvisited areas of search spaced rapidly. The TS selects one of the neighborhood functions based on the random probability during the search. In general, $N_1$ and $N_2$ take the partial CMCA $M$ as input and try to transform the rows in $M$ based on a different strategy. After each transformation, the algorithm evaluates the movement. If the search goes toward a better solution, the algorithm accepts that movement. Otherwise, it rejects it. In this case, the evaluation function $F(R)$ is a minimization function that is based on the number of covered $t-tuple$. The TS continue this process until all $t-tuple$ are covered.



The first neighborhood function $N_1$ receives the partial CMCA matrix $M$ and the number of rows $I_{rows}$ as an input. First, $N_1$ initializes a random row index $r$ and column index $c$. Then, the function starts with the $r^{th}$ row $R[r]$ and changes the value of $c^{th}$ column $R[c]$ in a way that randomly chooses a value from $V_{p_c}$ except the current value. After that, the changed row or a movement is checked for validity, as mentioned in Section \ref{InitializationSection}. The algorithm checks the validity for the only affected $t-tuples$. If the movement violates the constraints, it will be rejected, and the algorithm will try a new movement. Otherwise, it will be accepted. Later, the valid movement $R$ is evaluated. If the fitness value is less than or equal to the best solution, the movement $R$ is accepted and it will be the global best solution $R_{best}$. Otherwise, the row index $r$ is sequentially incremented, but the column index $c$ is fixed. This process is repeated for $I_{rows}$ iterations, as shown in \ref{Algorithm 5}. At some instances, the function may return zero converge. At this situation, the algorithm will stop and returns the complete CMCA $M$, lines (9 to 10). It is necessary to mention that the $N_1$ does not allow the search to return to the previous or explored regions which means that the global search and the evaluation function is based on the total number of uncovered $t-tuples$.

\setcounter{AlgoLine}{0}

\begin{algorithm}
\KwIn {partial M, number of rows $I_{rows}$}
\KwOut{optimized M}

$c \gets select\: random\: column\: 0\: to\: k$

$r \gets select\: random\: row\: 0\: to\: N$

$R_{best} \gets the\: numer\: of\: covered \;\; t-tuples$

$i \gets 0$

\While{$i < I_{rows}$}{

	$R \gets M[r]$
	
	$R[c] \gets randomly \: choose \: a \: value \: from \; V_{p_c} \; \setminus \; R[c]$
	
	\If{$ F(R) \leq F(R_{best}) $}{
	
		$R_{best} \gets f(R)$
		
		\If{$ F(R_{best}) = 0 $}{
		
			stop the algorithm
			
		}
		
		$i \gets i + 1$
		
	}
	
	$r \gets (r + 1) \mod N$
}

\caption{\label{Algorithm 5}The First Neighborhood Functions $N_1$}
\end{algorithm}


The second neighborhood function $N_2$ receives the partial CMCA matrix $M$ and the number of rows $I_{rows}$ to transform. Algorithm \ref{Algorithm 6} shows the steps of this function. First, $N_2$ selects a row index $r$ randomly except the row indices in the tabu list. Then, the $N_2$ starts with the $r^{th}$ row and replaces the randomly selected uncovered $t-tuple$ from the search space $S$ with the corresponded $t-tuple$ indices from R. The movement row is checked for validity, it will be accepted when it is valid. Otherwise, movement row will be rejected and the algorithm tries a new iteration for replacement movement. The movement $R$ is evaluated if the fitness value is less than or equal to the best solution obtained so far. At this stage, the movement $R$ is accepted and it will be the best local solution at $R_best$. Here, the current row index $r$ is added to the tabu list and it will be tabu for this function call. For the next iteration, a new random row is selected and the process is repeated for $I_{rows}$ iterations. The function may converge to zero; then the algorithm will stop and returns the complete CMCA $M$.

\setcounter{AlgoLine}{0}

\begin{algorithm}
\KwIn {partial M, number of rows $I_{rows}$}
\KwOut{optimized M}
$T \gets \{empty \: tabu \: list\}$

$r \gets 0$

$R_{best} \gets +\infty$

$i \gets 0$\;

\While{$i < I_{rows}$}{

	$r \gets select\: random\: row\: 0\: to\: N \setminus \{T\}$
	
	$R \gets M[r]$\;
	
	$R[(t-tuple)] \gets randomly \: choose \: uncovered \:\: t-tuple \: from \: S$
	
	\If{$ F(R) \leq F(R_{best}) $}{
	
		$R_{best} \gets f(R)$
		
		\If{$ F(R_{best}) = 0 $}{
		
			stop the algorithm
			
		}
		
		$i \gets i + 1$
		
	}
	
	$T \gets T \cup \{r\}$
}

\caption{\label{Algorithm 6}The Second Neighborhood Functions $N_2$}
\end{algorithm}

\section{Evaluation of results and discussion}\label{Evaluation}

This section describes the benchmarks used to evaluate the efficiency of the proposed method, the tools, and the approaches that the proposed method is compared against. The section also shows the settings of the experiments and the evaluation of experimental results. Finally, the section presents the effectiveness of the generated test suites through an empirical case study.


\subsection{Benchmarks}\label{Benchmarks}

In the experiments evaluating the proposed method, we use the well-known 35 standard constrained benchmarks that were originally used by \cite{Cohen2008}; later, details were introduced in \cite{Garvin2009} for evaluating different strategies. Five benchmarks out of these 35 benchmarks are models of real-world systems including an Apache server; an open-source http server; the GNU Compiler Collection (GCC); Bugzilla, a bug-tracking system; SPINV. a verified SPIN model checker; and SPINS, a simulator for SPIN. The remaining 30 benchmarks were inspired by the features of the five real-world models, such as the number of input parameters, the cardinality of each parameter and the complexity of the constraints \cite{Lin2016}.

\subsection{Approaches used for comparison}

We compare our proposed method with two state-of-the-art constrained approaches in the literature. These approaches are CASA \cite{Garvin2009} a Simulated Annealing (SA) based algorithm to generate constrained test suites and ACTS \cite{Yu-iinbin2013} a greedy algorithm to generate constrained test suites using the in-parameter-order (IPO) algorithm. CASA is an improved version of its base meta-heuristic algorithm SA for CIT \cite{Cohen2007}. CASA known as a benchmark strategy due to its efficiency of generating small sized CCA. CASA is a free open source project available for download\footnote{http://cse.unl.edu/\textasciitilde citportal/citportal/}. ACTS is an improvement of a greedy algorithm called IPOG \cite{Lei2007} and it consists of the IPO-family algorithms with three levels of optimization for constraint handling. In ACTS, the number of calling the constraint solver is reduced to reduce the runtime as a result. Therefore, ACTS is the best performing CCA generator as compared to the other approaches. ACTS is publicly available to download\footnote{https://csrc.nist.gov/projects/automated-combinatorial-testing-for-software/downloadable-tools}.


\subsection{Experimental Settings}

The proposed algorithm was implemented in the C\# language and compiled with .NET Core version 2.2 with enabled optimization options. We created three experimental groups based on the strengths of the interactions $t=2, 3, 4$. We set different values for the initial test suite (CMCA) size $N$, the iterations $I$ and the number of rows within the neighborhood functions $I_{rows}$. For CASA, which is written in C++, and ACTS, which is implemented in Java, we used the default settings. All the experimental sets were executed 30 times for all the tools except ACTS, a deterministic algorithm, where only one run produced the same result. The experiments were executed on a machine with a 2.6 GHz Intel Core i7 processor with 16 GB of memory in the macOS environment. To evaluate the efficiency of the proposed algorithm, we compared the proposed algorithm and two state-of-the-art approaches by using the benchmarks described in section \ref{Benchmarks} for generating 2-wise and 3-wise CCAs for all benchmarks and a 4-wise CCA for some of the benchmarks.

\subsection{Experimental results and discussion}

\begin{figure}
\centering
\includegraphics[width=1\textwidth]{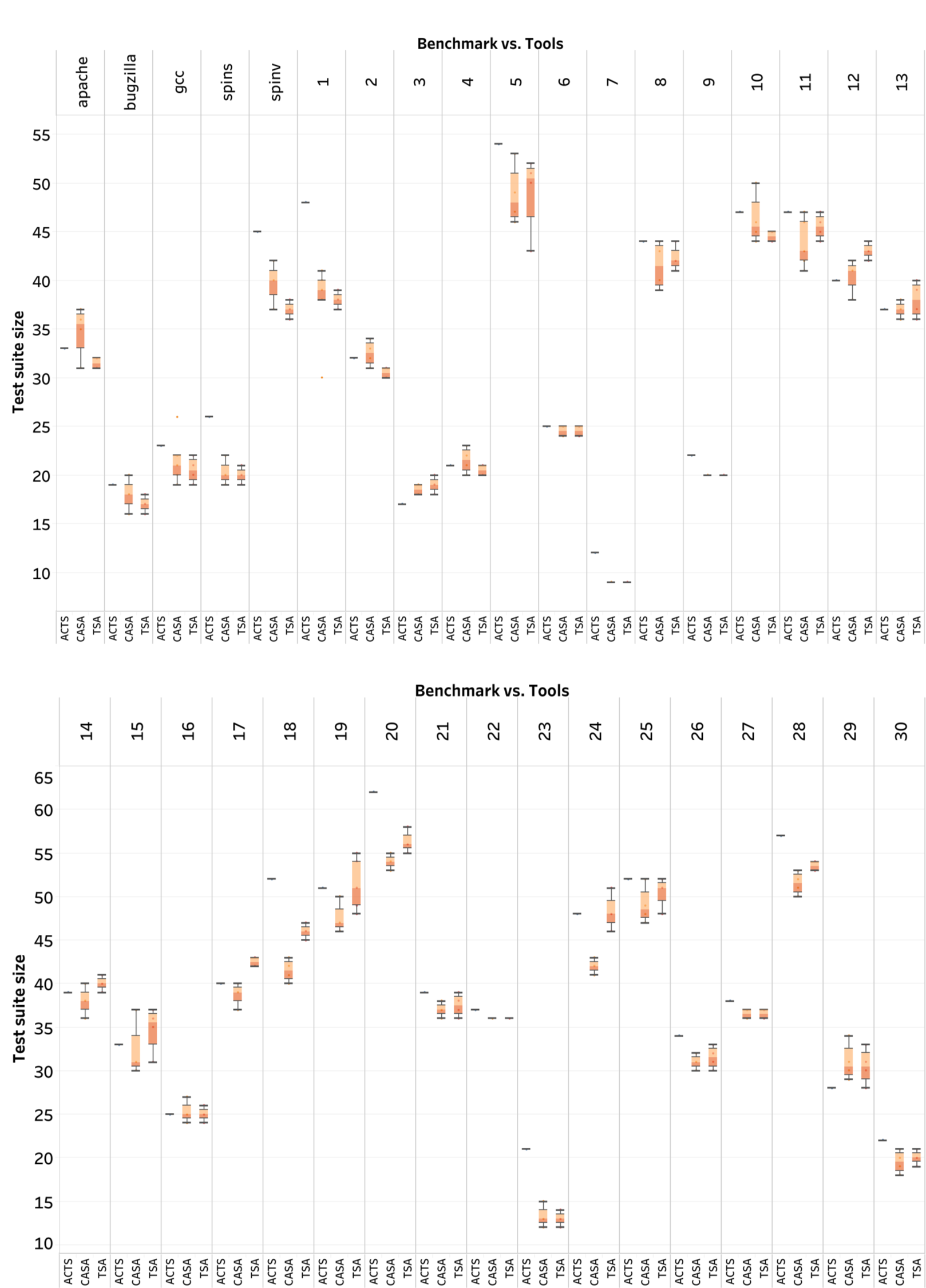}
\caption{Test suite size comparison between the proposed algorithm and competitors for strength 2 for all benchmarks}
\label{Figure2}
\end{figure}

The experimental evaluation of the efficiency in Figure \ref{Figure2} shows a comparison of the generated test suite sizes for the proposed strategy, which are labeled by TS algorithm (TSA), CASA and ACTS, with an interaction strength of $t=2$. Each column in the figure represents a benchmark that contains three box plots for each tool. It can be observed that the proposed method achieved almost the same results as CASA for most of the benchmarks except benchmarks 8, 11, 12, 14, 15, 17, 18, 19, 20, 24, 25, 28, and 30 in which CASA generated smaller test suites; however, for benchmarks 8, 11, 15, 25 and 30, both the TSA and CASA overlap with each other, which means that both tools produced similar results for many runs. However, for the Apache and SPINV benchmarks and benchmarks 1,2, 10, and 29, the TSA generated better results. The remaining benchmarks have very similar results. Regarding the comparison of the TSA with ACTS, which is a deterministic algorithm, both approaches always generated the same result, as it appeared as a line. The TSA outperformed the ACTS for all the benchmarks except benchmarks 3, 12, 15, 17 and 29. ACTS only generated smaller test suites for benchmarks 3 and 17. For benchmarks 14 and 29, ACTS achieved similar results. As observed from the results, we can clearly confirm that the TSA can compete and outperform its competitors when $t=2$.

\begin{figure}
\centering
\includegraphics[width=1\textwidth]{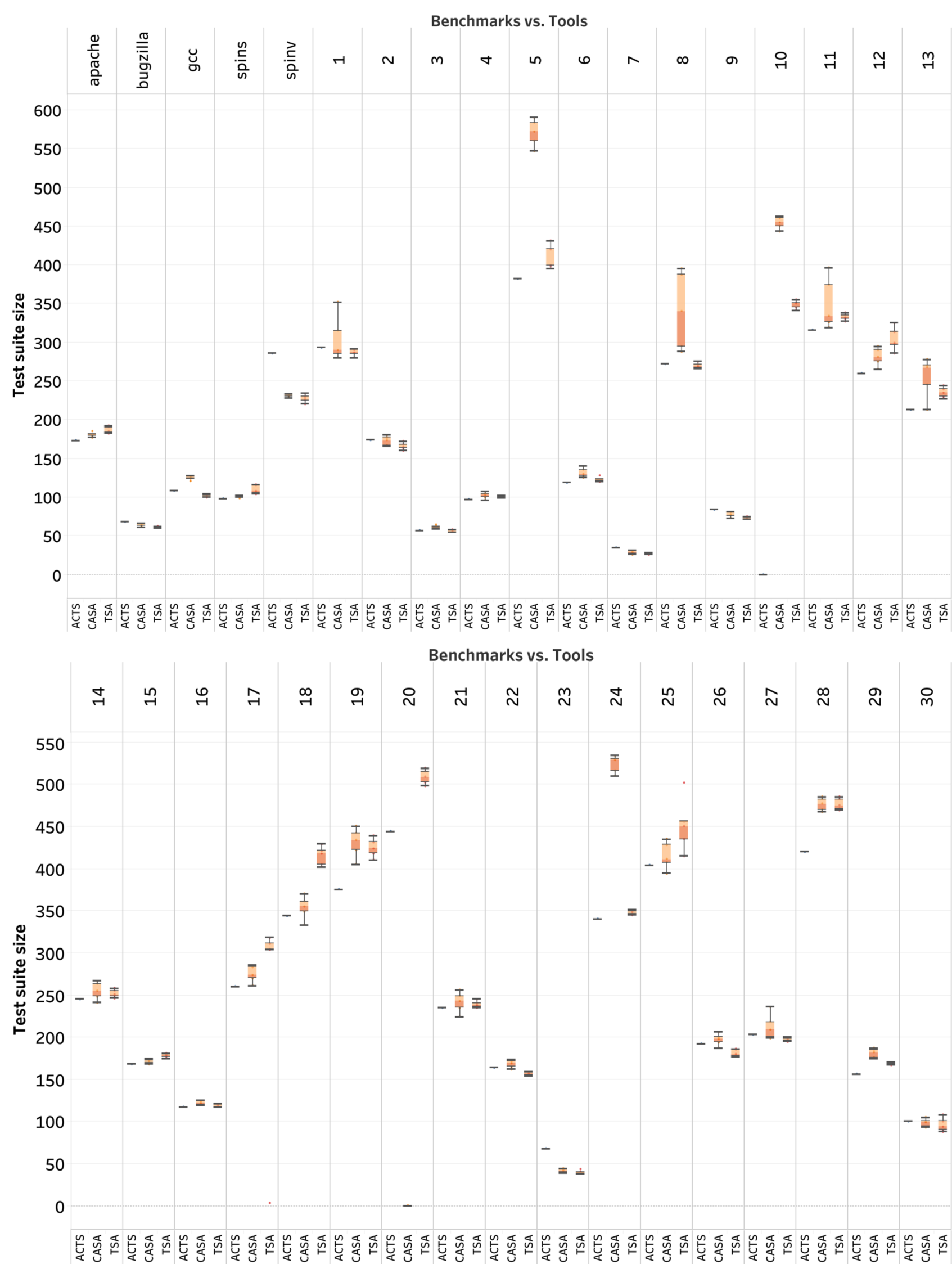}
\caption{Test suite size comparison between the proposed algorithm and competitors for strength 3 for all benchmarks}
\label{Figure3}
\end{figure}


Figure \ref{Figure3} shows test suite size comparisons for the second set of experiments with an interaction strength of $t=3$. Here, it is clear that the proposed method TSA could generate smaller test suites than CASA for 21 benchmarks out of 35 except for the Apache and SPINS benchmarks and benchmarks 12, 13, 14, 15, 17, 18 and 25. In the case of benchmarks 1, 7, 9, and 28, the TSA produced similar results. For benchmarks 5, 8, 10, 24, and 26, the TSA produced far smaller results than CASA. It is noticeable here that CASA could not generate a result for benchmark 20. If we compare the test suite sizes generated by the TSA and ACTS, the TSA achieved better results than ACTS for 15 benchmarks. For benchmarks 16 and 21, both tools generated the same results. For the reaming benchmarks, the TSA generated larger test suites. For example, for benchmarks 19, 20 and 28, ACTS notably produced smaller test suites than both the TSA and CASA because such benchmarks have more than 150 parameters, which leads to a large search space. ACTS is a deterministic algorithm that is fast, but both the other tools, the TSA and CASA, are stochastic algorithms that may generate smaller test suites if the initial parameter and interaction are increased; however, the TSA and CASA will require much more time to cover all the valid $t-tuples$. Based on this fact, for the benchmarks that have a smaller number of parameters and a smaller search space, the TSA outperformed ACTS when the strength was $t=3$. However, as mentioned before, when the interaction strength was 2, the TSA outperformed ACTS for most of the benchmarks, including the benchmarks with large inputs, since the search space increases exponentially when the interaction strength increases. We should mention here that ACTS could not generate a test suite for benchmark 10.

\begin{figure}[b]
\centering
\includegraphics[width=1\textwidth]{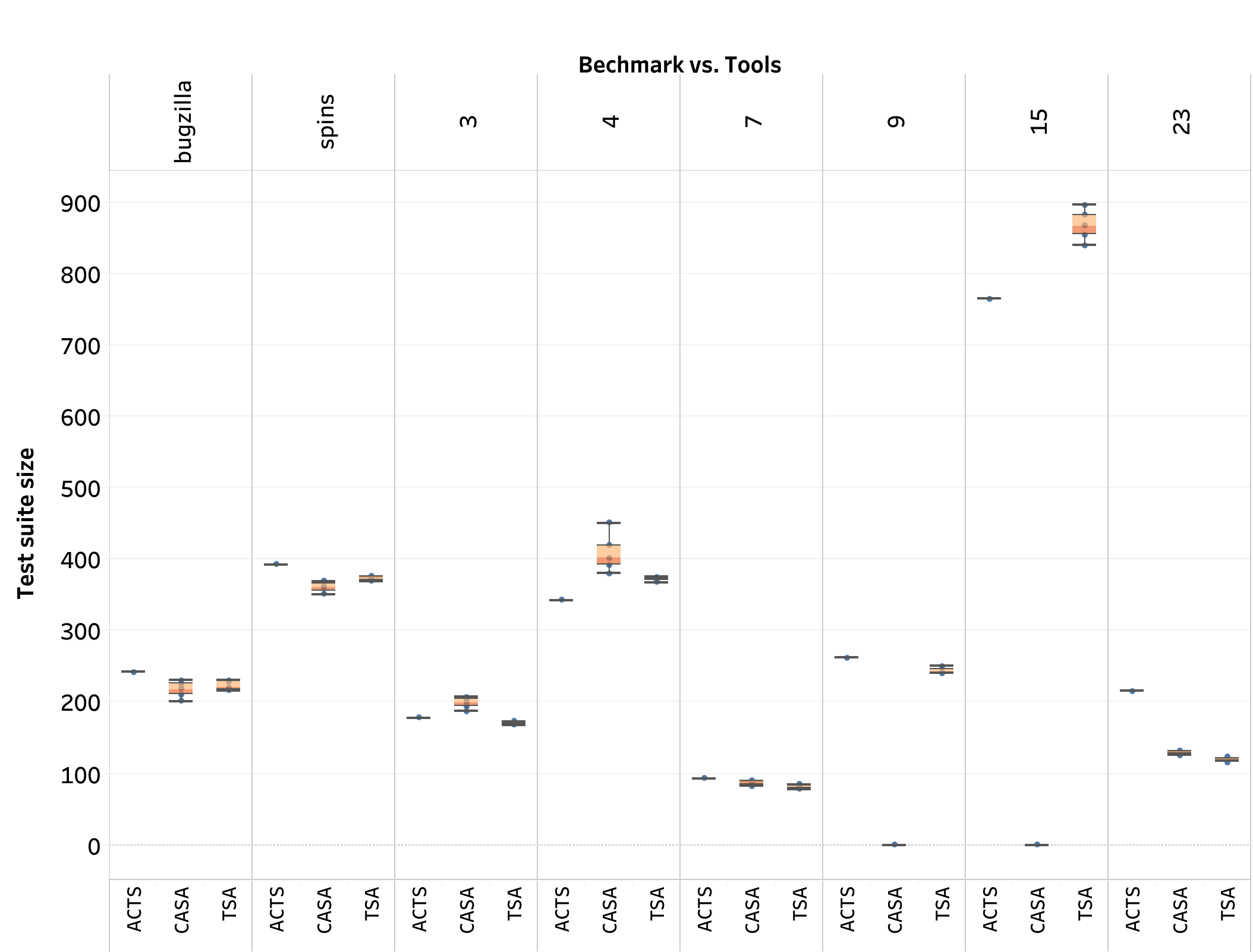}
\caption{Test suite size comparison between the proposed algorithm and competitors for strength 4 for some benchmarks}
\label{Figure4}
\end{figure}

Finally, Figure \ref{Figure4} shows the results of the last set of experiments for an interaction strength of $t=4$. It is clear that the proposed TSA could only generate test suites for eight benchmarks with fewer than 80 parameters. The results showed that the average test suite sizes generated by the TSA achieved better results than CASA for benchmarks 3, 4, 7 and 23. Except for the Bugzilla and SPINS benchmarks in which it generated larger test suites with small differences, CASA could not generate a result for benchmarks 9 and 15. Comparing the results of the TSA with those of ACTS, it is clear that the TSA generated smaller test suites for 6 benchmarks except for benchmarks 4 and 15, in which the TSA generated impressive results.


\subsection{Effectiveness evaluation through a case study}

\begin{figure}
\centering
\includegraphics[width=0.7\textwidth]{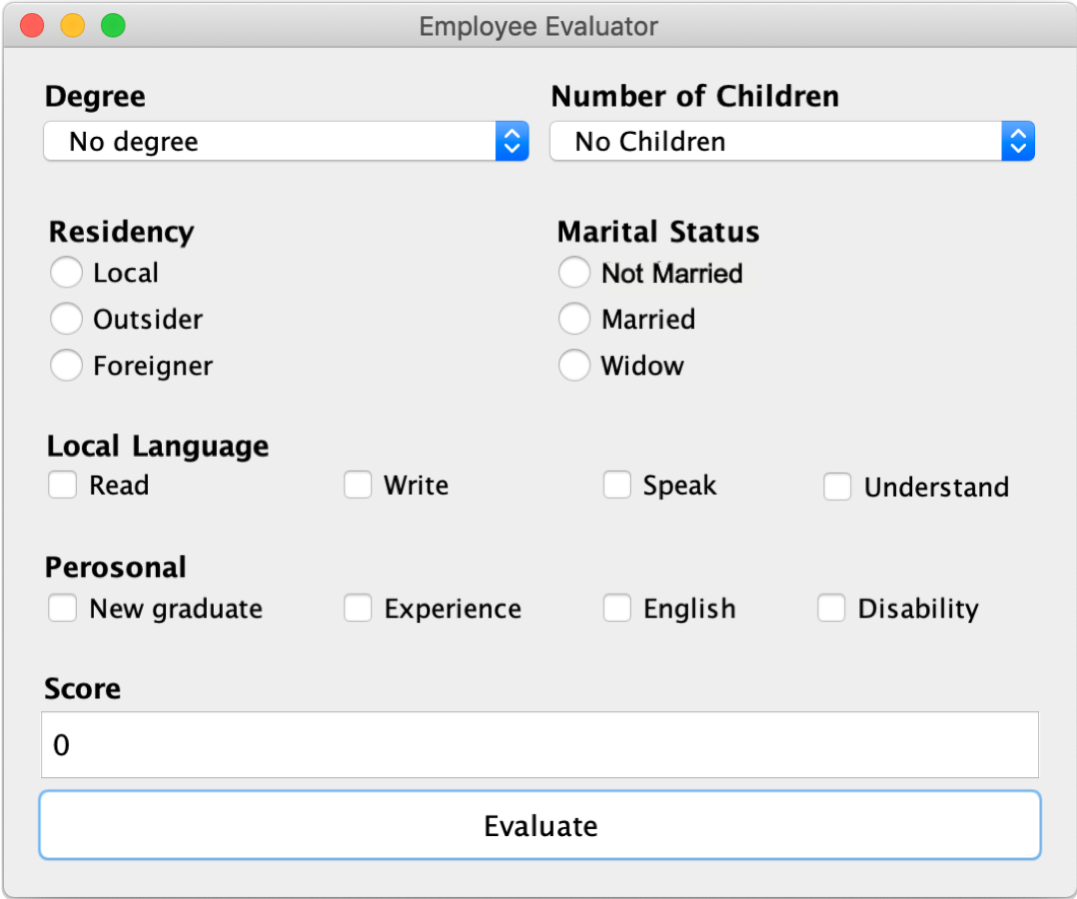}
\caption{The GUI program of the case study}
\label{Figure5}
\end{figure}

To show the effectiveness and quality of the generated test suites, we chose a special program as a real-world subject for a case study. The program was adopted from \cite{Ahmed2015}; however, we modified the program by adding constraints to its logic. This program evaluates and ranks employees by assigning them a weighted score based on their personal and academic information. The score is used to calculate their monthly salary. This program has a nontrivial codebase and has different input configurations. As shown in Figure \ref{Figure5}, the program can interact through a graphical user interface (GUI) that consists of different components. Each component represents a configuration as an input parameter. Each input parameter has several different values. For example, the input parameter ``Residency'' has three different values: ``Local'', ``Outsider'' and ``Foreigner''. Additionally, the input parameter ``Experience'' is either ``False'' or ``True''. Table \ref{Table 4} shows a complete list of the input parameters and their values for the case study.

It is clear from Table \ref{Table 4} that the input parameters of the SUT can be represented by one parameter with seven values, one parameter with six values, two parameters with three values and eight parameters with two values. As mentioned before, there are constraints in the configurations of the SUT that observed from the domain knowledge of the application such as when the value of parameter ``Marital Status'' is ``Not Married'' the parameter ``Number of Children'' should be ``No children'', and when the value of parameter ``Degree'' is ``No degree'', the parameter ``English'' should be unchecked ``False''. The identified constraints are represented as a set of forbidden tuples $C$, as shown below. Then, $C$ is encoded to a processable set and passed to constraint handling algorithm to convert it to a set of BFTs. Finally, the processed $C$ is passed to the test generation algorithm for validity check.

\begin{align*} 
C &= \{     \{(Marital \; Status, Not \: Married), (Number \; of \; children, 1)\},\\
& \qquad \! \{(Marital \; Status, Not \: Married), (Number \; of \; children, 2)\},\\
& \qquad \! \{(Marital \; Status, Not \: Married), (Number \; of \; children, 3)\},\\
& \qquad \! \{(Marital \; Status, Not \: Married), (Number \; of \; children, 4)\},\\
& \qquad \! \{(Marital \; Status, Not \: Married), (Number \; of \; children, More \; than \; 4)\},\\
& \qquad \! \{(Degree, No \; Degree), (English, True)\}\}
\end{align*}

Accordingly, the input parameter model of the SUT can be represented in the form of CMCA notation as $CMCA(N,t,7^1 6^1 3^2 2^8, C)$ where $N$ is the size of the test suite, $t$ is the interaction strength and $C$ is the constraint set. To generate the exhaustive testing for all parameter combinations, we need to test 96,768 possibilities. However, with the CCIT techniques, the size of the test suite can be reduced dramatically. Table \ref{Table 5} shows the different test suites sizes of the SUT generated by our TSA for different interaction strengths.

\begin{table}
\caption{\label{Table 4}The input parameters with their values of the GUI
program}

\raggedright{}%
\begin{tabular*}{1\textwidth}{@{\extracolsep{\fill}}ll>{\raggedright}p{0.5\textwidth}c}
\toprule 
{\footnotesize{}\#} & {\footnotesize{}Parameter} & \multicolumn{1}{l}{{\footnotesize{}Values}} & {\footnotesize{}Cardinality}\tabularnewline
\midrule
\midrule 
{\footnotesize{}1} & {\footnotesize{}Degree} & {\footnotesize{}No \: degree, Primary, Secondary, Diploma, Bachelor,
Master, Doctorate} & {\footnotesize{}7}\tabularnewline
\midrule 
{\footnotesize{}2} & {\footnotesize{}Number of children} & {\footnotesize{}No children, 1, 2, 3, 4, More than 4} & {\footnotesize{}6}\tabularnewline
\midrule 
{\footnotesize{}3} & {\footnotesize{}Residency} & {\footnotesize{}Local, Outsider, Foreigner} & {\footnotesize{}3}\tabularnewline
\midrule 
{\footnotesize{}4} & {\footnotesize{}Marital Status} & {\footnotesize{} Not Married, Married, Widow} & {\footnotesize{}3}\tabularnewline
\midrule 
{\footnotesize{}5} & {\footnotesize{}Read} & {\footnotesize{}False, True} & {\footnotesize{}2}\tabularnewline
\midrule 
{\footnotesize{}6} & {\footnotesize{}Write} & {\footnotesize{}False, True} & {\footnotesize{}2}\tabularnewline
\midrule 
{\footnotesize{}7} & {\footnotesize{}Speak} & {\footnotesize{}False, True} & {\footnotesize{}2}\tabularnewline
\midrule 
{\footnotesize{}8} & {\footnotesize{}Understand} & {\footnotesize{}False, True} & {\footnotesize{}2}\tabularnewline
\midrule 
{\footnotesize{}9} & {\footnotesize{}New graduate} & {\footnotesize{}False, True} & {\footnotesize{}2}\tabularnewline
\midrule 
{\footnotesize{}10} & {\footnotesize{}Experience} & {\footnotesize{}False, True} & {\footnotesize{}2}\tabularnewline
\midrule 
{\footnotesize{}11} & {\footnotesize{}English} & {\footnotesize{}False, True} & {\footnotesize{}2}\tabularnewline
\midrule 
{\footnotesize{}12} & {\footnotesize{}Disability} & {\footnotesize{}False, True} & {\footnotesize{}2}\tabularnewline
\bottomrule
\end{tabular*}
\end{table}

\begin{table}
\caption{\label{Table 5}The SUT test suites for $t = 1 to 6$}

\raggedright{}\centering%
\begin{tabular}{cc}
\toprule 
Interaction strength (t) & Size of test suite\tabularnewline
\midrule
\midrule 
2 & 44\tabularnewline
\midrule 
3 & 140\tabularnewline
\midrule 
4 & 378\tabularnewline
\midrule 
5 & 870\tabularnewline
\midrule 
6 & 2139\tabularnewline
\bottomrule
\end{tabular}
\end{table}


To measure the quality and effectiveness of the generated test suites, we used the mutation testing technique to assess the quality of the generated test suites. Here, different types of mutants (faults) are injected into the SUT using Pitest \footnote{http://pitest.org/}. Pitest is a mutation testing tool that first creates various types of faults within the source code. For each fault, the tool creates a version or a copy of the injected source code with faults. Then, it applies the generated test suites to all the injected versions of the SUT using the unit testing strategy. Finally, it generates a mutation testing report that contains information on how many mutations were detected (killed) and not detected (not killed) or covered and not covered. Here, there are two important advantages of using mutation testing. First, it verifies how the components in the source code of the SUT interact with each other and how they affect the flow of data from the input to the output. This is known as test case behavior. The second advantage is the test suite reduction. From the generated report of the mutation testing, sometimes, similar test cases can be detected if they have the same behavior of killing and not killing identical injected mutations. Thus, mutation testing verifies and improves the final test suite size generated by the proposed method by removing test cases that have the same behavior as the injected mutations.

\begin{figure}
\centering
\includegraphics[width=1\textwidth]{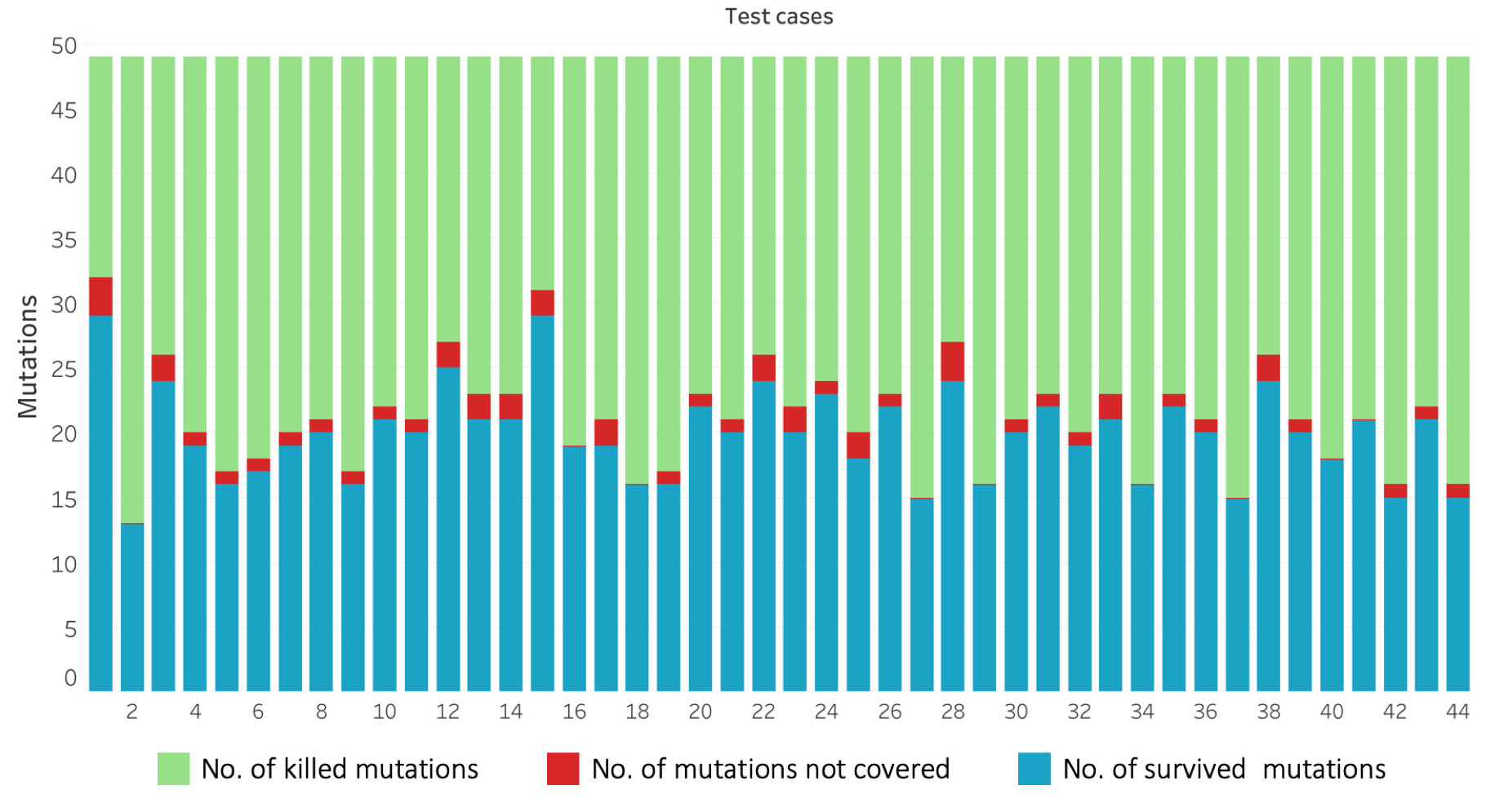}
\caption{The Analysis of mutation testing report on the case study for the test suite when $t=2$}
\label{Figure6} 
\end{figure}

For the verification step, we encoded the behavior of the injected mutations for all test cases as a vector. This vector will be the pattern of behaviors of the test cases. Then, we calculated the hash code for all vectors to find similar test cases that have the same hash code. In Table \ref{Table 5}, when $t=2$, the proposed method generated a test suite with a  size of 44. We performed mutation testing on this test suite. The Pitest tool generated 49 mutations that were injected into the source code of the SUT. As a result, 49 versions of the SUT source code were produced. The test suite was applied to the mutation-injected SUT. Figure \ref{Figure6} shows the generated report of the 49 injected mutations against the 44 test cases. Each stacked bar represents a reaction of the test case against all 49 mutations. The blue part from the stacked bar is the number of mutations that were not detected (not killed). This means that the injected mutation did not affect the output result. The test case did not cover the red part of the mutations. The green part is the number of detected or killed mutations; thus, the injected mutations had a direct effect on the output result. The report indicated that overall the test cases detected 96\% of the injected mutations, or 47 mutations out of 49.

\begin{figure}
\centering
\includegraphics[width=1\textwidth]{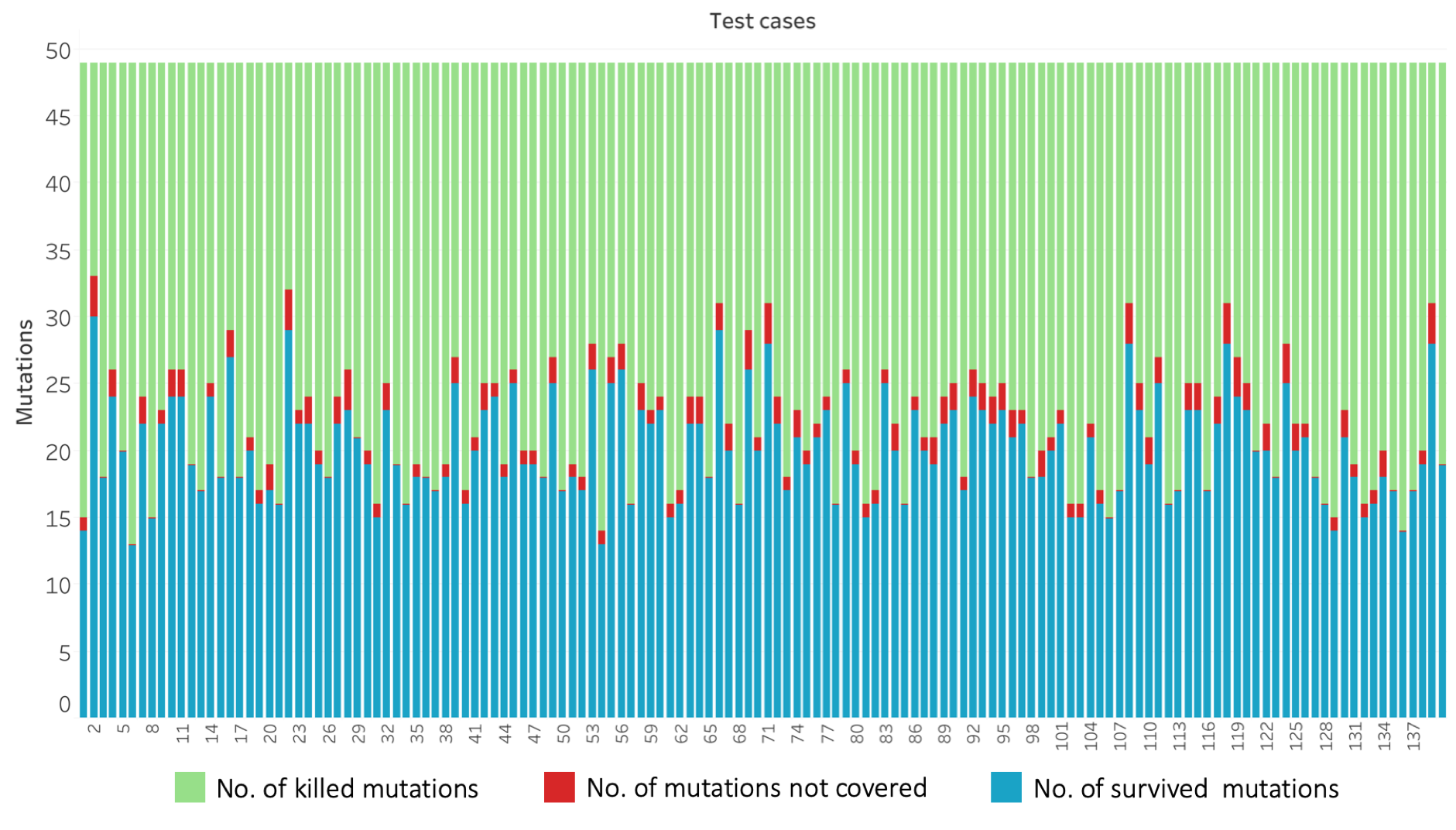}
\caption{The Analysis of mutation testing report on the case study for the test suite when $t=3$}
\label{Figure7}
\end{figure}

We checked the generated report for the test cases that have similar behaviors. The reactions of each test case against the injected mutations are grouped and encoded as a vector of behavior patterns, and then the vector is hashed. There were no redundant test cases in the hash codes. As can be noted in Figure \ref{Figure6}, many of the test cases have the same stacked bars. For example, test case groups (4,7), (5, 9, 19), (10,21), (18, 29, 34), (20, 31), (22, 38), (27, 37), (36, 39), and (42 and 44) have similar stacked bars. However, this does not guarantee that these test cases have similar behaviors because they are only similar in the number of killed, not killed and not covered mutations. Note that in the test case reactions against the injected mutations, it is clear that their behavior pattern vectors do not match. Similarly, when $t=3$, the proposed method generated 140 test cases. For the mutation testing, 49 similar mutations were injected into the SUT and tested against the new test suite. Figure \ref{Figure7} shows an analysis of the mutation testing report. The analysis shows the reaction of the test cases against the 49 injected mutations. The report was checked for redundant test cases in terms of behavioral similarity using the same strategy as that used for the first test suite. We found only one test case with redundant behavior. Test cases 45 and 83, rather than having the same stacked bars, exhibit the same pattern of behavior and have the same hash code. Additionally, many test cases have the same stacked bar pattern, for example, most notably, test case groups (10,11), (47, 48), (63,64), and (102,103). However, these test cases are not considered similar test cases because they only share the same number of killed, not killed and not covered mutations.

It is noticeable that the test cases cannot detect all the injected mutations or faults individually. However, overall the test cases could cover the entire source code and detect all the injected faults except for two faults that the test suite with a strength of 2 failed to detect. From the mutation testing reports, 47 injected faults out of 49 were detected, which means that 95.9\% of the faults were detected by the test suites with strengths of 2 and 3. We also applied mutation testing for strengths of 4 and higher and obtained the same fault detection rate as that for a strength of 2. We noticed that these faults were still undetectable when we increased the interaction strength. This was due to the nature of the internal logic of the case study.


\section{Threats to Validity}\label{Threats}

As with other studies, the experiments in this study suffer from a few threats to validity such as:

\begin{enumerate}
\item Choosing the 35 standard benchmarks as a model of SUT for efficiency evaluation is a source of threats to validity since they may not represent every scenario for regular real-world system models.

\item The proposed strategy and one of its competitors CASA are using a stochastic algorithm. This means that in each execution, they produce different results. 

\item One more threat that encountered the proposed method in the experiments is the input iterations variable. The larger iterations produce better results (smaller test suites) but with longer execution time. 

\item Only one ideal example is selected as a case study to evaluate the effectiveness of generated test suites, which may not fit for all the scenarios.
\end{enumerate}

During the experiments, we tried to follow several well-known practices and steps to mitigate and overcome these threats. For the first threat, the five benchmarks from the 35 benchmarks are real-world systems; these five benchmarks inspire the rest. Therefore, analysis and evaluations may be generalized. To mitigate the second threat, we executed the strategies for 30 runs for each benchmark. To reduce the third threat, we tuned the control parameters and designed different experiment configurations for the benchmarks. To overcome the last threat, many other real-world system models can be used as a case study to show the effectiveness of test suites. Since, in the current example, we tried to inject real faults and use real program for evaluation. However, this is just a case study and several other case studies may exist that can be used within our strategy. Even with the use of many case studies, generalizing the results is challenging because they are case-specific results and they may change based on the application. However, we aim to show that our strategy can detect the interaction faults for other applications since it can detect the current one.

\section{Related Work}\label{RelatedWork}

CIT has been broadly studied in the last decade, and many approaches have been proposed to generate CIT test suites. The proposed methods in the literature are either constrained or unconstrained. Nie and Leung \cite{Nie2011} presented an excellent survey on unconstrained CIT approaches, concepts, and applications. Kuhn \textit{et al.} \cite{Kuhn2013} broadly covered many practical aspects of unconstrained CIT. Ahmed \textit{et al.} \cite{Ahmed2017} presented an intensive and systematic literature survey of constrained CIT, addressing the current generation strategies, tools, and applications. From the mentioned studies, it can be observed that there are three categories of generation approaches that can be classified based on generation techniques: greedy algorithms, metaheuristic search algorithms, and mathematical or algebraic methods. Additionally, there is another generation approach that uses constraint satisfiability solvers for constraint handling, which is usually adopted with the two most popular greedy and metaheuristic search algorithms.

Greedy algorithms are the most widely used approach for test suite generation. There are two generation strategies in the case of a greedy algorithm: the one-parameter-at-a-time (OPTAT) strategy and the one-test-at-a-time (OTAT) strategy. An example of the OPTAT strategy is an in-parameter-order (IPO) algorithm that starts by generating a test suite with only $t$ parameters; then, the algorithm extends the test suite by adding one parameter in each iteration. There are two types of test suite extensions: horizontal and vertical extensions. The above process is repeated until all valid $t-tuples$ are covered. The IPO algorithm was originally proposed by Lei and Tai \cite{Lei1998} for generating 2-wise test suites, but it was generalized for t-wise test suites as an in-parameter-order-general (IPOG) algorithm by \cite{Lei2007}. The combinatorial test generation tool (ACTS) presented by \cite{Yu2013} accounts for most of the IPO algorithms. Within the OTAT approach, the test suite is constructed by generating one test in each iteration until all valid $t-tuples$ are covered. The purpose of using such a strategy is to cover valid $t-tuples$ as much as possible in each iteration. A well-known example of this strategy is the automatic efficient test generator (AETG) \cite{Cohen1997}, which was the first greedy algorithm to use the OTAT strategy. Similarly, there are many AETG variations, including \cite{Tung2000} and \cite{Bryce2007}.


Many metaheuristic algorithms have been proposed in the literature for the construction of CIT, e.g., the genetic algorithm (GA) \cite{McCaffrey2009}, simulating annealing (SA) \cite{Cohen2003}, and particle swarm optimization (PSO) \cite{Ahmed2012}. The strategies in the literature employ a similar strategy for generating the test suite. The strategy usually starts by constructing a partial or incomplete test suite and then applies modifications or transformation until it covers all the valid $t-tuples$. In each iteration, the algorithm moves toward unexplored regions so that it tries to cover all possible missed valid $t-tuples$. Hernandez \textit{et al.} \cite{Gonzalez-Hernandez2010org} also used the TS algorithm to generate CIT test suites. In that study, the authors used a mixture of three neighborhood functions that were selected based on probability and focused on CCIT test suites. We also proposed a new algorithm to handle the constraints. Another approach that utilized the TS algorithm was the TCA algorithm proposed by Lin \textit{et al.} \cite{Lin2016}. Their contribution was using a two-mode framework that combined the modes of a greedy algorithm and random generation. Based on probability, the algorithm switched between the two modes. In the greedy mode, the TCA employed a TS to cover the valid $t-tuples$ that allowed generating an entire row from the search space. However, the authors only showed the experimental results for interaction strengths of 2 and 3 and used the same SAT-based constraint solver as that used with CASA; the scalability of such an SAT solver is challenging for higher interaction strengths due to the process of encoding  the test case and the initial constraint into a boolean formula. However, our approach enables test suite generation for an interaction strength of 4 for some of the same benchmarks, and we also proposed a more scalable constraint-handling strategy.

Another proposed TS algorithm is the Covering Array by Tabu Search (CATS) \cite{Galinier2017}. The main contribution of this work was the penalty based local search algorithm in which the algorithm allows illegal solutions that violate the constraints during the search. The illegal solutions help the search space to be more explored thoroughly, but in the end, the invalid $t-tuples$ are penalized and removed from the final test suite. They reported the results of well-known 35 benchmarks for only interaction strength 2 that have minimum generation time so far with the same quality as TCA. However, for today's high-configurable systems, higher interaction strength is essential to detect the faults with higher parameter combinations as we reported the results of the same benchmarks for higher interaction strength.

As for the support of CCIT, Covering Arrays by Simulated Annealing (CASA) is another tool that proposed by Garvin \textit{et al.} \cite{Garvin2011} that improved the previously implemented SA algorithm for generating CCIT test suites that originally proposed by Cohen \textit{et al.} \cite{Cohen2007}. CASA has shown the substantial results for $t=2$ for the well-known 35 benchmarks. However, the performance and the size of generated test suites degraded for $t > 2$.

In the literature, a few methods have been proposed for handling constraints. Cohen \textit{et al.} \cite{Cohen2008} used an SAT solver for an AETG-like algorithm called mAETG to provide support for constraint handling. Garvin \textit{et al.} \cite{Garvin2011} improved the SA algorithm by adding the constraint SAT solver for constraint-handling support. In parallel, Garvin \textit{et al.} also modified the original SA algorithm for better integration with the constraint SAT solver and used them in the CASA tool. Although constraint SAT solvers  are useful, there are different methods that can be used for constraint handling. For example, PICT \cite{Czerwonka2008} is another tool that proposed the forbidden tuple strategy as a complement to the greedy algorithm. PICT first generates all the forbidden tuples; then, these tuples are used to check the validity during the test generation \cite{Yamada2016}. Yu \textit{et al.} proposed a strategy called minimum invalid tuples (MITs) for constraint handling for ACTS \cite{Yu2014}. The strategy generates all the forbidden tuples at the beginning, and then the tuples are used for a test case validation. Later, the authors improved and generalized this approach in \cite{Yu2015} with an approach called minimum forbidden tuples (MFTs) that generated the forbidden tuples only when required using an on-demand strategy called the necessary forbidden tuple (NFT) strategy. The study concluded that the MFTs with the on-demand strategy outperformed the MIT and PICT approaches.

In our proposed strategy, we integrated a new constraint handler with an improved TS algorithm. We also introduced a new way of validity checking. Specifically, when the TS algorithm transforms a solution to move toward another solution, the only affected parameters are checked, not the whole parameter combinations in the new solution. This helps to reduce the time spent in validity checking.


\section{Conclusion}\label{Conclusion}

In this paper, we presented an approach for CCIT that utilizes a TS within a scheme for constraint handling called the BFT algorithm. The main contribution of this paper is the use of a forbidden tuple with a metaheuristic algorithm that utilizes two neighborhood functions $N_1$ and $N_2$ to explore the search space. Another contribution of the paper is the improvements in the validity checking; for each move, it only checks the $t-tuples$ that contain affected parameters. To evaluate the proposed algorithm, we used 35 well-known constrained benchmarks with interaction strengths of $t=2, 3$, and $4$. The evaluation results showed that our strategy could compete with the CASA tool when $t=2$ and could generate smaller test suites than CASA when $t=3$ for most of the benchmarks. The results also showed that our strategy outperformed ACTS for most of the benchmarks when $t=2$ and $t=3$. For some specific benchmarks, our proposed strategy achieved better results when $t=4$. We also carried out an empirical case study to evaluate the effectiveness of the generates test suites using mutation testing. The evaluation results showed that the test suites generated by our strategy could detect most of the mutants injected into the SUT. This result indicated the effectiveness of the generated test suites.


The main future challenge of this research is the investigation of results with high interaction strengths in the presence of constraints. We intend to investigate search space and time optimization schemes. Our approach and the other tools were  unable to generate 4-wise test suites for all the benchmarks due to memory limitations. However, in contrast to the other tools, our strategy could generate 4-wise test suites for benchmarks with less than 100 parameters. There are some techniques in the literature that might help reduce the memory size of the data structure, such as compression techniques or using a different strategy for storing the $t-tuple$ sets. There is also a possibility to optimize the search time by implementing our approach in a parallel computing environment such as a compute unified device architecture (CUDA) \cite{Ploskas2016}, which is a framework based on C/C++ and an NVIDIA graphics processing unit (GPU), or GPU programming with MATLAB\cite{Schmidt2018}. There is also a possibility to produce better test suites by fine-tuning the control parameters of the search algorithm and neighborhood functions. For example, increasing the main iteration $I$ in the algorithm leads to an increased calling of the neighborhood functions in each TS call, which leads to better solutions but takes more time. On the other hand, increasing the number of rows $I_{rows}$ to modify each neighborhood function may also lead to a better solution by more extensively exploring the search space regions.

\section*{References}

\bibliographystyle{model1-num-names}
\bibliography{sample}

\end{document}